\documentclass{gGAF2e}
\usepackage{graphicx}
\usepackage{amsmath}
\usepackage{fmtcount}
\usepackage{ctable}
\usepackage{setspace}
\usepackage{multirow}
\usepackage[normalem]{ulem} 
\usepackage{color}
\begin{document}
\jvol{00} \jnum{00} \jyear{2014} \jmonth{February}

\title{ASYMPTOTIC SOLUTIONS FOR MEAN-FIELD SLAB DYNAMOS}
\author{YAMENG JI, 
LAURA COLE$^{\ast}$\thanks{$^\ast$Corresponding author. Email: L.Cole@ncl.ac.uk\vspace{6pt}}, PAUL BUSHBY 
and ANVAR SHUKUROV\\[6pt]
School of Mathematics and Statistics, Newcastle University, Newcastle upon Tyne, NE1 7RU, UK\\[6pt]
\received{\today}}
\markboth{Y.~Ji \textit{et al.}}{Asymptotic solutions for mean-field slab dynamos}

\renewcommand*{\Phi}{\varPhi}
\renewcommand*{\Psi}{\varPsi}
\renewcommand*{\Gamma}{\varGamma}
\renewcommand*{\Theta}{\varTheta}

  \newcommand{\aj}{{Astron.\ J.}}
  \newcommand{\aap}{{Astron.\ Astrophys.}}
  \newcommand{\apj}{{Astrophys.\ J.}}
  \newcommand{\apjl}{{Astrophys.\ J.\ Lett.}} 
  \newcommand{\araa}{{Ann.\ Rev.\ Astron.\ Astrophys.}}
  \newcommand{\mnras}{{MNRAS}}
  \newcommand{\rmp}{{Rev.\ Mod.\ Phys.}}
  \newcommand{\jfm}{{J.\ Fluid Mech.}} 
  \newcommand{\sovast}{{Sov.\ Astron.}} 
  \newcommand{\apss}{{Astrophys.\ Space Sci.}}

\newcommand\bgreek[1]{ \mathchoice
    {\hbox{\boldmath$\displaystyle{#1}$\unboldmath}}%
    {\hbox{\boldmath$\textstyle{#1}$\unboldmath}}%
    {\hbox{\boldmath$\scriptstyle{#1}$\unboldmath}}%
    {\hbox{\boldmath$\scriptscriptstyle{#1}$\unboldmath}}}
    
\newcommand{\la}{\mathrel{\mathchoice {\vcenter{\offinterlineskip\halign{\hfil
$\displaystyle##$\hfil\cr<\cr\sim\cr}}}
{\vcenter{\offinterlineskip\halign{\hfil$\textstyle##$\hfil\cr<\cr\sim\cr}}}
{\vcenter{\offinterlineskip\halign{\hfil$\scriptstyle##$\hfil\cr<\cr\sim\cr}}}
{\vcenter{\offinterlineskip\halign{\hfil$\scriptscriptstyle##$\hfil\cr<\cr\sim\cr}}}}}

\newcommand{\ga}{\mathrel{\mathchoice {\vcenter{\offinterlineskip\halign{\hfil
$\displaystyle##$\hfil\cr>\cr\sim\cr}}}
{\vcenter{\offinterlineskip\halign{\hfil$\textstyle##$\hfil\cr>\cr\sim\cr}}}
{\vcenter{\offinterlineskip\halign{\hfil$\scriptstyle##$\hfil\cr>\cr\sim\cr}}}
{\vcenter{\offinterlineskip\halign{\hfil$\scriptscriptstyle##$\hfil\cr>\cr\sim\cr}}}}}

%
  %
  \newcommand{\cm}{\,{\rm cm}}
  \newcommand{\m}{\,{\rm m}}
  \newcommand{\cmcube}{\,{\rm cm^{-3}}}
  \newcommand{\dyn}{\,{\rm dyn}}
  \newcommand{\erg}{\,{\rm erg}}
  \newcommand{\g}{\,{\rm g}}
  \newcommand{\Jy}{\,{\rm Jy}}
  \newcommand{\Jyb}{\,{\rm Jy/beam}}
  \newcommand{\km}{\,{\rm km}}
  \newcommand{\kms}{\,{\rm km\,s^{-1}}}
  \newcommand{\mJy}{\,{\rm mJy}}
  \newcommand{\mJyb}{\,{\rm mJy/beam}}
  \newcommand{\K}{\,{\rm K}}
  \newcommand{\kpc}{\,{\rm kpc}}
  \newcommand{\pc}{\,{\rm pc}}
  \newcommand{\Mpc}{\,{\rm Mpc}}
  \newcommand{\Myr}{\,{\rm Myr}}
  \newcommand{\Gyr}{\,{\rm Gyr}}
  \newcommand{\mG}{\,{\rm mG}}
  \newcommand{\mkG}{\,\mu{\rm G}}
  \newcommand{\nG}{\,{\rm nG}}
  \newcommand{\MHz}{\, {\rm MHz}}
  \newcommand{\Msol}{\,{\rm M_{\sun}}}
  \newcommand{\p}{\,{\rm pc}}
  \newcommand{\radm}{\,{\rm rad\,m^{-2}}}
  \newcommand{\s}{\,{\rm s}}
  \newcommand{\yr}{\,{\rm yr}}   

\newcommand{\const}{\mathrm{const}}	
\newcommand{\crit}{_\mathrm{cr}}		
\newcommand{\inex}{\mathrm{d\llap{$^-$}}}
\newcommand{\ex}{\mathrm{e}}
\newcommand{\ix}{\mathrm{i}}
\newcommand{\de}{\mathrm{d}}
\newcommand{\eql}[2]{\begin{equation} #1 \label {#2} \end{equation}}
\newcommand{\eq}[1]{\begin{equation} #1  \end{equation}}
\newcommand{\eqn}[1]{\begin{eqnarray} #1  \end{eqnarray}}
\newcommand{\tr}[3]{{#1}_{#2}^{#3}}
\newcommand{\half}{\frac{1}{2}}

\maketitle

\begin{abstract}
We discuss asymptotic solutions of the kinematic $\alpha\omega$-dynamo in a thin disc (slab)
surrounded by an electric insulator. Focusing upon the strong dynamo regime, 
in which the dynamo number $D$ satisfies $|D|\gg1$, 
we resolve uncertainties in the earlier treatments and conclude that some of the 
simplifications that have been made in previous studies are questionable. 
Having abandoned these simplifications, we show, by comparing numerical solutions 
with complementary asymptotic results obtained for $|D|\gg1$ and $|D|\ll1$, that the asymptotic solutions 
give a reasonably accurate description of the dynamo even far beyond their formal ranges 
of applicability. Indeed, our results suggest a simple analytical expression for the 
growth rate of the mean magnetic field that remains accurate 
across the wide range of values for $D$ that are typical of spiral galaxies and accretion discs. 
Finally, we analyse the role of various terms in the governing equations to clarify the 
fine details of the dynamo process.
In particular, in the case of the radial magnetic field equation we have shown that the $\alpha\,\upartial B_\phi/\upartial z$ term (where $B_\phi$ is the azimuthal magnetic field, $\alpha$ is the mean-field dynamo coefficient, and $z$ is measured across the slab), which is neglected in some of the earlier asymptotic studies, is essential for the dynamo as it drives a flux of magnetic energy away from the dynamo region, towards the surface of the slab.

\begin{keywords}
Mean-field dynamos; Asymptotic solutions; Numerical solutions; Galaxies and accretion discs
\end{keywords}

\end{abstract}


\section{Introduction}\label{intro}
Mean-field slab (or thin disc) dynamos can be represented by a very simple system of equations, with $z$ (the distance across the slab, parallel to the rotation axis) as the only spatial variable. In the case of astrophysical discs, it is usually appropriate to work with the $\alpha\omega$-dynamo equations, in which it is assumed that the mean inductive effects of the rotational shear (the $\omega$-effect) are significantly stronger than those due to the effects of helical turbulence (the $\alpha$-effect). In systems of this type, the behaviour of the dynamo depends upon the spatial distribution of the $\alpha$-effect across the slab $\alpha(z)$, as well as the imposed boundary conditions and the assumed differential rotation profile. However, once these features have been specified, the $\alpha\omega$-dynamo only has a single control parameter, the dynamo number $D$, which measures the efficiency of the source terms in the dynamo equations relative to magnetic diffusion. Given the importance of the slab dynamos for applications (e.g., galactic and accretion discs), they have attracted a significant amount of attention \citep[e.g., ][]{M78,P79,RTZS79,RSS88,BBMSS96,S07,SS08}. The simple structure of their eigensolutions, with a discrete spectrum of real eigenvalues under conditions typical of spiral galaxies and accretion discs, 
makes this system attractively accessible to analysis. 

Exact solutions are known for kinematic $\alpha\omega$ slab dynamos with a few artificial 
(discontinuous) forms of $\alpha(z)$. Examples include a piecewise constant distribution 
$\alpha(z)=\alpha_0$ for $z>0$, $-\alpha_0$ for $z<0$ and $0$ for $z=0$ \citep{P79}, 
and a distribution consisting of two delta functions, $\alpha(z)=\alpha_0[\delta(z-z_0)-\delta(z+z_0)]$,
with $\alpha_0=\const$  \citep{M78,RTZS79}. These functional forms for $\alpha(z)$ produce both oscillatory and non-oscillatory solutions. However, numerical solutions with continuous $\alpha(z)$ are non-oscillatory for negative dynamo numbers of a moderate magnitude \citep{RST80,SL90}. A deeper insight into the nature of the eigensolutions and their dependence upon the distribution of $\alpha(z)$ has been provided by approximate and asymptotic solutions.

\citet{IRSF81} developed boundary-layer asymptotics for an $\alpha\omega$-dynamo in a slab for $|D|\gg1$. Adopting a definition for the dynamo number that is consistent with that given below (see equation~\eqref{Ddef}), these authors were able to show that the growth rate of the leading mode, which has quadrupolar parity, has the form
\begin{equation}\label{eq:growthrate}
\gamma \simeq \bar{\gamma}_0 \left|\alpha'(0) D \right|^{1/2}\,,
\end{equation}
where $\alpha'(0)$ is the value of $\de\alpha/\de z$ at $z=0$ and $\bar{\gamma}_0$ is a constant of order unity. 
In their study, this constant was determined to be $\bar{\gamma}_0\approx0.3$
from fitting this form to a numerical solution. These authors used further simplifications 
to reduce the problem to a second-order ordinary differential equation which still is not 
amenable to analytic solution. We discuss and assess these simplifications below
(see also Appendix~\ref{SC}, where we show, in particular, that the simplified asymptotic boundary value problem 
considered by these authors in fact has only trivial
quadrupolar solutions for $D<0$). 
Following \citet{IRSF81}, \citet{S95} similarly simplified the $\alpha\omega$-dynamo equations 
to obtain, from a WKB asymptotic solution, $\gamma\simeq  \left|\alpha'(0) D \right|^{1/2}$. 
It remains unclear how these asymptotic solutions are related to each other, how they compare 
to the numerical ones, and what their ranges of applicability are. 

In this paper, we revisit the asymptotic solutions of the kinematic $\alpha\omega$-dynamo in a slab. 
We discuss the various approximations involved and resolve some earlier uncertainties to present what is arguably a definitive asymptotic analysis for $|D|\gg1$. These asymptotics are then compared with both numerical
solutions and analytical results from perturbation theory (formally applicable for $|D|\ll1$). We demonstrate 
that both asymptotic solutions, for $|D|\gg1$ and $|D|\ll1$, remain remarkably accurate far beyond their formal
ranges of applicability. Thus, solutions obtained and discussed here offer a firm foundation for further qualitative
studies and applications to galactic and accretion-disc dynamos. Having presented the dynamo equations in section~\ref{DE}, we discuss in section~\ref{ASDE} the form of their asymptotic solutions in the kinematic regime, i.e., when the back-reaction of magnetic fields on the velocity field is still negligible and the 
magnetic field strength grows exponentially with time. Numerical solutions
for the slab $\alpha\omega$-dynamo surrounded by a vacuum are presented in section~\ref{NS}, 
and then compared to the asymptotic solutions for both $|D|\gg1$ and $|D|\ll1$. This clarifies 
the applicability of various assumptions employed in deriving the asymptotic solutions and 
allows us to establish their final form. The implications of our results are discussed in 
section~\ref{Disc}. Various simplifications and their pitfalls are discussed in Appendix~\ref{SC}.

\section{Disc dynamos}\label{DE}
The evolution of a large-scale magnetic field $\bm{B}$ in an electrically conducting fluid
is governed by the mean-field dynamo equation \citep[e.g.,][]{M78,KR80} 
\begin{equation} \label{eq:kinematic}
\frac{\upartial \bm{B}}{\upartial t} = \bm{\nabla} \times ( \bm{V} \times  \bm{B}) + 
\bm{\nabla} \times (\alpha  \bm{B})  - \eta \bm{\nabla}\times\bm{\nabla}\times\bm{B},
\end{equation}
where $\alpha$ represents the mean-field $\alpha$-effect (production of the mean magnetic field 
by a mirror-asymmetric random flow), $\bm{V}$ is the large-scale fluid velocity and 
$\eta$ is the magnetic diffusivity (usually dominated by its turbulent part) which is assumed, 
for simplicity, to be independent of position. In the framework of the kinematic dynamo theory, 
$\bm{V}$ and $\alpha$ are prescribed functions of position, and are thus independent of $\bm{B}$. 

We introduce cylindrical polar coordinates $(r, \phi, z)$ with the origin at the disc centre and the $z$-axis 
perpendicular to the disc (or slab) plane. The disc is assumed to have a horizontal size $L$ and thickness $2h$. 
We restrict our attention to axisymmetric solutions of equation~\eqref{eq:kinematic} and assume that  
$L \gg h$, meaning that the disc is thin. The advantage of considering the thin disc limit is that 
the terms involving radial derivatives (and those in $\phi$) are much smaller than those with derivatives in $z$, and can therefore be neglected in the lowest approximation \citep{RSS88}. In most 
astrophysical applications, the large-scale velocity field is dominated by an overall rotation
and, thus, is predominantly azimuthal. Since $\upartial V/\upartial z\approx0$ at $z\approx0$ 
from symmetry considerations, the large-scale velocity field in a thin layer is practically
independent of $z$. We can therefore assume that the large-scale velocity field takes the form 
$\bm{V}(r) = (0,r \varOmega(r),0)$.As $\varOmega$ decreases with $r$ in most cases of interest, we can also assume that $\de\varOmega/\de r<0$. The rotational velocity shear rate is given by 
$G(r)= r\, \de\varOmega / \de r$, which will be assumed to be constant in this study, 
$G(r)=G_0$, with $G_0<0$. 
In a thin layer, the evolution equation for $B_z$ decouples from the equations for $B_r$ and $B_{\phi}$, so only the latter two equations must be considered in the subsequent analysis. Once these equations have been solved, it is straightforward to calculate $B_z$ using the fact that $\bm{\nabla}\cdot\bm{B}=0$ \citep{RSS88}. 

Dimensionless variables, denoted with a circumflex, can be conveniently defined as follows:
\[
t =\frac{h^2}{\eta} \widehat{t}\,, 
\quad 
r = L \widehat{r}\,, 
\quad 
z = h \widehat{z}\,, 
\quad 
B_r = B_0 \widehat{B_r}\,, 
\quad 
B_\phi = \frac{B_0 \eta}{h \alpha_0} \widehat{B_\phi}\,, 
\quad 
\alpha = \alpha_0 \widehat \alpha\,, 
\]
where $B_0$ and $\alpha_0$ are representative values of the magnetic field and $\alpha$-effect, respectively. Throughout the remainder of this paper, all variables will be dimensionless unless stated otherwise. So, to simplify notation, we drop the circumflex in what follows. Having introduced these scalings, we are left with the following dimensionless equations for the magnetic field components $B_r$ and $B_{\phi}$:
\begin{eqnarray}
\label{eq:inductionr}
\frac{\upartial {B}_r }{\upartial t} &=& - \frac{\upartial}{\upartial z}(\alpha B_{\phi}) 
+ \frac{\upartial^2 B_r}{\upartial z^2}\,, \\
\label{eq:inductionphi}
\frac{\upartial {B}_\phi}{\upartial t} &=& D B_r 
+ \frac{\upartial^2 B_{\phi} }{\upartial z^2}\,,
\end{eqnarray}
where 
\begin{equation}\label{Ddef}
D= \frac{\alpha_0 G_0 h^3}{\eta^2}
\end{equation}
is the dynamo number; $|D|$ must exceed a certain critical value $|D\crit|$ to allow 
magnetic field to grow.

As the velocity field is independent of $\bm{B}$, solutions have the form
\[
\bm{B}(z,t) = \bm{b}(z) \ex^{\gamma t}\,,
\]
leading to
\begin{eqnarray}
\label{eq:br}\gamma b_r &=& - \frac{\de}{\de z}(\alpha b_{\phi}) 
	+ \frac{{\mathrm{d}^2} b_r }{\mathrm{d} z^2} \, , \\
\label{eq:bphi}\gamma b_{\phi} &=& D b_r + \frac{{\mathrm{d}^2} b_{\phi} }{\mathrm{d} z^2}\,.
\end{eqnarray}
From the overall symmetry of the problem, $\alpha$ is an odd function of $z$, which implies that $\alpha(z)=-\alpha(-z)$ with $\alpha(z)>0$ for $z>0$ \citep{RSS88}. As a consequence of this the solutions are either quadrupolar [with $b_{r,\phi}(-z)=b_{r,\phi}(z)$ and $b_z(-z)=-b_z(z)$] or dipolar [with $b_{r,\phi}(-z)=-b_{r,\phi}(z)$ and $b_z(-z)=b_z(z)$]. Earlier results demonstrate that quadrupolar modes dominate strongly in a thin layer surrounded by an electric insulator (vacuum) \citep{M78,RSS88}.  Therefore, we consider the range $0\leq z\leq1$ and apply the following boundary conditions at the disc mid-plane that respect the quadrupolar symmetry:
\begin{equation}\label{scond}
\frac{\de b_r}{\de z} = \frac{\de  b_\phi}{\de z} =0
\quad
\mbox{at}\ z=0.
\end{equation}
At the surface of the layer, $z=1$, we adopt the vacuum boundary conditions \citep{M78,RSS88}, 
\begin{equation}\label{bcond}
b_r  = b_\phi = 0
\quad
\mbox{at}\ z=1\,.
\end{equation}
An eigenvalue problem has now been formulated, with $\gamma$ the eigenvalue, $(b_r,b_\phi)$ 
the eigenfunction and $D$ the control parameter. The solution obviously depends upon the spatial distribution of $\alpha(z)$. However, this dependence is not strong. For example, regardless of the choice for $\alpha(z)$, it is rare for the critical dynamo number to lie outside of the narrow range  $-13\leq D\crit\leq-4$ \citep{RSS88}, even with a discontinuous distribution for the $\alpha$-effect.

\section{Boundary layer produced by strong dynamo action
}\label{ASDE}
In the outer parts of spiral galaxies, $\left| D \right| \sim 10$, whilst values of a few hundred 
are typical of their central regions \citep[e.g.,][]{S07}. 
It is therefore reasonable to look for asymptotic solutions for large dynamo number, 
$\left|D\right| \gg 1$. As discussed here, for large $|D|$ a boundary layer develops
at $z\approx0$.

In this analysis, we consider smooth functional forms for the $\alpha$-effect. Since $\alpha(z)$ is an odd function, it can be expanded as $\alpha(z) \approx \alpha_1 z + \alpha_3 z^3+...$ in the vicinity of $z=0$ (where $\alpha_1$ and $\alpha_3$ are dimensionless constants of order unity). The special case of $\alpha_1=0$ is discussed briefly below. However, unless stated otherwise, we shall assume that $\alpha_1$ is strictly positive throughout this section. In this case, we can further assume (without loss of generality) that $\alpha_1=1$ because we are free to choose an appropriate scaling for the $\alpha$-effect when the equations are made dimensionless. So, at leading order, $\alpha(z) \approx z$ in the vicinity of $z=0$. 

\par We introduce the scaled variable,
\[
s = |D|^{k} z\,,
\]
and assume the following scalings within the boundary layer:
\[
\frac{\mathrm{d}}{\mathrm{d}z}=|D|^{k}\frac{\mathrm{d}}{\mathrm{d}s}\,,
\qquad
b_r = \left| D \right|^n R(s)\,,
\qquad 
b_{\phi} = \Phi(s)\,,
\qquad
\gamma = \Gamma_0 |D|^m\,,
\]
where $k$, $m$ and $n$ must be determined. Having written the term responsible for the 
$\alpha$-effect in equation~(\ref{eq:br}) as
\begin{equation} \label{eq:alphab}
\frac{\mathrm{d}}{\mathrm{d} z}(\alpha b_{\phi}) = \frac{\mathrm{d} \alpha}{\mathrm{d} z}  b_{\phi} + \alpha \frac{\mathrm{d} b_{\phi}}{\mathrm{d} z}\,,
\end{equation}
we note that, within the boundary layer,
\[
\frac{\mathrm{d} \alpha}{\mathrm{d} z}  b_{\phi} \approx \Phi(s)=\mathrm{O}(1)\,,
\]
and
\[
\alpha \frac{\mathrm{d} b_{\phi}}{\mathrm{d} z} \approx z  \frac{\mathrm{d} b_{\phi}}{\mathrm{d} z}  
= s \frac{\mathrm{d}\Phi}{\mathrm{d}s}=\mathrm{O}(1)\,.
\]
Thus, the two terms in $\de(\alpha b_\phi)/\de z$ are of 
the same order of magnitude in $\left|D\right|$ 
and neither of them can be neglected. It is straightforward to show that this is also the case when $\alpha_1 =0$
(i.e., $\alpha=\alpha_3z^3+\ldots$). Although quadrupolar solutions tend to be preferred in this disc dynamo system, it can be shown that these terms would also be of the same order of magnitude if dipolar boundary conditions had been adopted, so this is a very general result.

\par For these quadrupolar solutions, all of the terms in equations~\eqref{eq:br} and \eqref{eq:bphi} are of the same order of magnitude in $|D|$ if 
\[
k=1/4,
\quad
m=1/2,
\quad
n=-1/2.
\]
Then the governing equations reduce to
\begin{eqnarray}
\label{eq:blf_1}
\Gamma_0 R &=& -\frac{\mathrm{d}}{\mathrm{d}s}(s\Phi) + \frac{\mathrm{d}^2R}{\mathrm{d}s^2}\,,\\
\label{eq:blg_1}
\Gamma_0 \Phi &=& \mbox{sign}(D)\, R + \frac{\mathrm{d}^2\Phi}{\mathrm{d}s^2}\,,\\
\label{bsc}
\frac{\de R}{\de s}&(0)& = \frac{\de \Phi}{\de s}(0) =0\,,
\quad	
|R|, |\Phi|\rightarrow0\ \mbox{as}\ s\rightarrow\infty\,.							
\end{eqnarray}
These boundary-layer equations are no simpler to solve than the original ones. 
This fact has prompted earlier authors to neglect $s\,\de\Phi/\de s$ in the first equation. 
It is now clear that this is unacceptable in any physically justifiable case. 
More details regarding possible simplifications can be found in Appendix~\ref{SC}.

Useful results can be obtained even from the form of the asymptotic solutions now clarified.
In particular, we have shown that $\displaystyle{b_{\phi}/b_r}$ and $\gamma$ are both 
$\displaystyle{\mathrm{O}( \left| D \right|^{1/2})}$ quantities in this asymptotic regime 
[cf.\ equation~\eqref{eq:growthrate}], 
whilst the magnetic field eigenfunctions vary over a characteristic spatial scale 
$\displaystyle{\mathrm{O}( \left| D \right|^{-1/4})}$. These quantitative predictions can be tested 
numerically and used in applications. \citet{IRSF81} obtained similar scalings but their further analysis involved a strong 
assumption that $|R(s)|\propto |\Phi(s)|$. This assumption is not supported by the numerical 
solutions reported in section~\ref{CNAS}. 

Similar boundary-layer equations can be derived in the case $\alpha_1=0$. If the first non-zero term of 
the Taylor-series expansion for $\alpha(z)$ is  $\alpha_p z^p$ (with $\alpha_p>0$), 
all of the terms in the governing 
equations contribute at the same order in $\left| D\right|$ if 
$\displaystyle{m=2/(p+3), \; n=-(p+1)/(p+3)}$ and $\displaystyle{k=1/(p+3)}$.

We should make one final comment before proceeding. As stressed by \citet{S95}, 
the asymptotic solutions discussed here are of intermediate nature as they apply when, 
on the one hand, $|D|\gg1$ but, on the other hand, 
when $|D|$ is not
too large. For $D\la-300$, the leading eigenvalue can be complex and dipolar modes can
become dominant. Under these circumstances, these solutions will no longer be valid. 
However, these parameter regimes are arguably less relevant for galactic applications.

\section{Numerical solutions}\label{NS}\label{CNAS}
In order to clarify the nature of the asymptotic solutions and to assess the validity 
of various approximations, we solved numerically for the leading eigensolution
of equations~\eqref{eq:inductionr} and \eqref{eq:inductionphi}, adopting the quadrupolar symmetry
conditions \eqref{scond} and vacuum boundary conditions \eqref{bcond} written for $B_r$ and $B_\phi$. 
We also assume here that 
\[
\alpha=\sin\pi z\,,
\]
so that $\alpha_1=\pi$. In addition, following \citet{IRSF81}, we shall obtain an accurate value of the
constant $\bar{\gamma}_0$ in the asymptotic expression \eqref{eq:growthrate}  for the eigenvalue. We used 
a 
fourth-order Runge--Kutta time stepping scheme, adopting a second-order finite difference approximation for the spatial derivatives (including the boundary conditions). All calculations made use of $N=100$ mesh points in $0\leq z\leq1$. This numerical resolution is more than adequate
for the values of $D$ that have been considered. 

\setcounter{table}{0}
\begin{table}
\tbl{The growth rate $\gamma$ of the magnetic field, for various values of the dynamo number $D$, 
for the leading quadrupole mode. The upper table shows our results, which were obtained using $N=100$ mesh points in $0\leq z\leq1$, whilst the lower part of the table shows the results with $N=10$ from \citet{RST80}. The entries in 
the
lower part of the table should be compared with the bold entries in the upper part of the table.}
{\begin{tabular}{cccccccc}\toprule
\multicolumn{8}{c}{$N=100$}\\
$D$ & $-5$   		&  $-8$ 			&   $-10$ & $-12$  &$-15$  &  $-20$  & $-25$ \\
$\gamma$ & $\mathbf{-0.50}$  	& $\mathbf{0.00}$ & $\mathbf{0.27}$ & $0.52$& $0.85$&$\mathbf{1.33}$ &$1.73$ \\[6pt]
$D$ 						& $-30$  			& $-35$  & $-40$ & $-50$  & $-60$  & $-70 $ & $-80$  \\
$\gamma $ & $2.09$ & $2.41$ & $2.70$ &$\mathbf{3.20}$ &$3.64$ &$4.01$ & $4.34$ \\ [6pt]
$D$ & $-90$  		& $-100$ & $-120$ & $-140$ & $-160$& $-180$ & $-190$ \\ 
$\gamma $ &$4.63$ &$\mathbf{4.88}$ &$5.30$ & $5.62$ & $5.85$ & $6.00$& $6.04$ \\[6pt]
$D$ & $-200$ 		& $-205$ & $-210$ & $-220$ & $-250$ & $-270$ & $-300$ \\
$\gamma $ & $6.07$& $6.08$& $6.07$ & $6.06$ &$5.88$ & $5.62$& $\mathbf{4.87}$ \\
\colrule
\multicolumn{8}{c}{$N=10$}\\
$D$ & $-5$ 			& $-8$ & $-10$ & $-20$ & $-50$ & $-100$ & $-300$ \\
$\gamma$ 	& $-0.5$& $-0.0$& $0.3$ & $1.4$ &$3.4$ & $5.2$& $5.6$\\ 
\botrule
\end{tabular}}
\label{gammaD}
\end{table}

\setcounter{figure}{0}
\begin{figure}
\begin{center}
\includegraphics[width=0.7\textwidth]{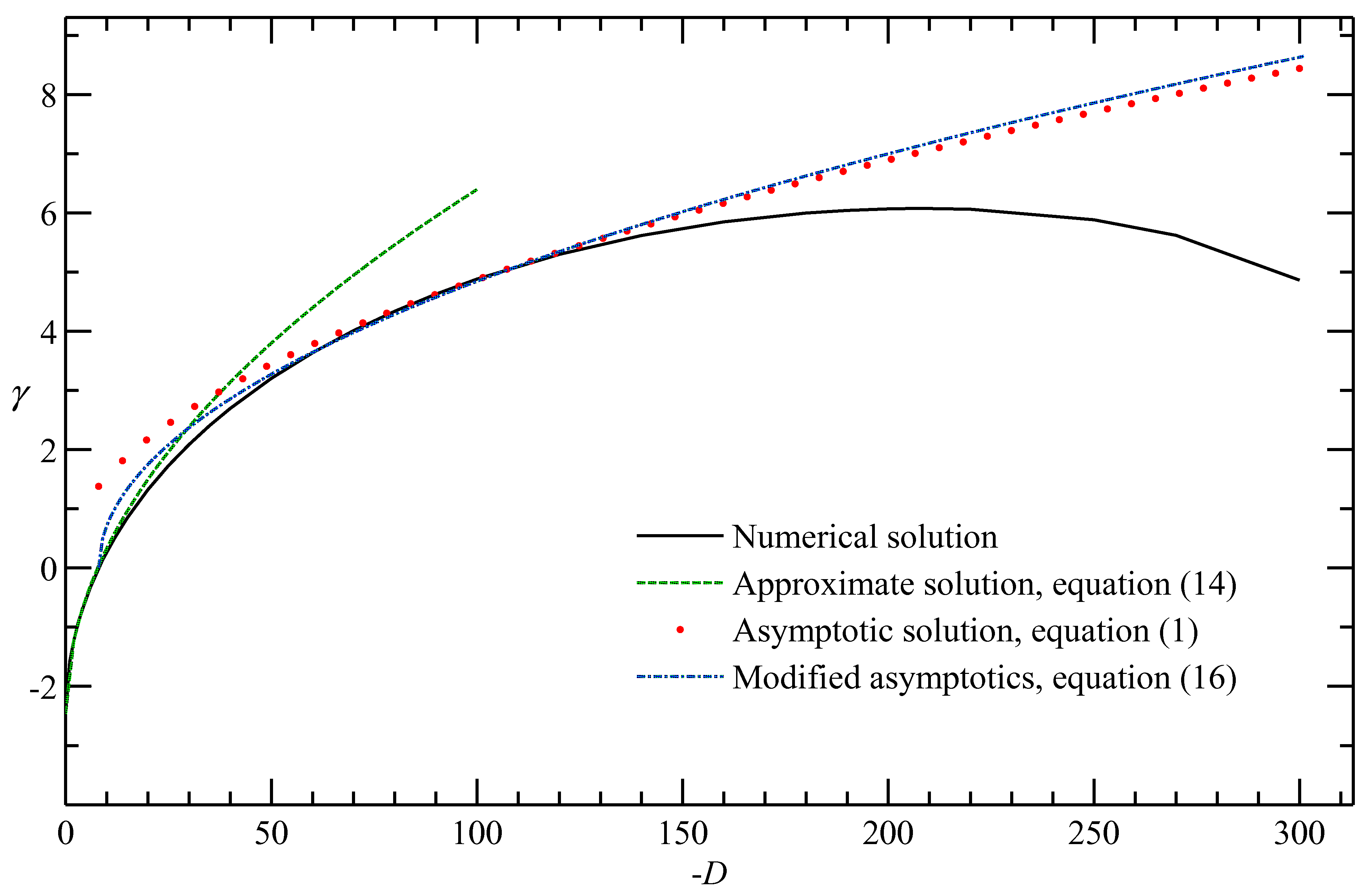}
\end{center}
\begin{center}
\caption{\label{fig:ln_gamma_vs_D}
The dependence of the growth rate of the magnetic field on the dynamo number, with numerical results 
being compared to various approximations. The lines show the numerical solution with $\alpha=\sin\pi z$ 
(black, solid),
the asymptotic solution \eqref{eq:growthrate} for $|D|\gg1$ with $\bar{\gamma_0} = 0.275$ 
(red, dotted), 
the perturbation solution \eqref{Dll1} for $|D|\ll1$ 
(green, dashed),
and the modified asymptotic form \eqref{g0275} for $|D|\gg1$ with $\alpha'(0)=\pi$ 
(blue, dash-dotted).}
\end{center}
\end{figure}

Focusing entirely upon negative values of the dynamo number ($-5 \ge D \ge -300$), 
we present in Table~\ref{gammaD} the dynamo growth rate $\gamma$ for a range of $D$, and
compare it with the earlier results of \citet{RST80} (which were obtained at a lower numerical resolution).  For all dynamo numbers in this range, the solutions are (as expected) non-oscillatory, with the magnetic field growing for $D<D\crit\approx-8$.  Despite significant differences in the numerical resolution, the results compare favourably 
with those of \citet{RST80} for small values of $|D|$ \citep[see also][]{SL90}. At larger values of $|D|$, we observe a local maximum in the growth rate, with $\gamma$ reaching a peak value at $D\approx-205$. The growth rate of the non-oscillatory quadrupolar mode then decreases for larger values of $|D|$.  

\subsection{The growth rate of the magnetic field}\label{GRMF}
The dependence of the growth rate of the magnetic field upon the dynamo number is shown in figure~\ref{fig:ln_gamma_vs_D}.  For small values of $|D|$, these results compare very favourably to the corresponding perturbation solution (for this choice of $\alpha=\sin \pi z$) of \citet{SS08},
\begin{equation}\label{Dll1}
\gamma = -\tfrac14\pi^2+ \tfrac12\sqrt{-\pi D} +\mathrm{O}(|D|^{3/2})\,,
\qquad
\alpha=\sin\pi z\,,
\qquad
|D|\ll1\,.
\end{equation}
This result is only weakly dependent upon the choice of functional form for $\alpha(z)$. For example, for $\alpha=z$, we have for the leading quadrupolar mode
\[
\gamma = -\tfrac14\pi^2+ \sqrt{-\tfrac12 D} + \mathrm{O}(|D|^{3/2})\,,
\qquad
\alpha=z\,,
\qquad
|D|\ll1\,,
\]

where the factor in front of $\sqrt{-D}$ differs from that in equation~\eqref{Dll1} by 25\%,
leading to $D\crit\approx-12$ instead of $-8$. This difference is arguably negligible 
given the uncertainties and scatter of galactic parameters as well as other idealisations
adopted in the dynamo model. Since the asymptotics for $|D|\gg1$
are even less sensitive to the form of $\alpha(z)$ away from the boundary layer near $z=0$, we conclude
that the results obtained with $\alpha=\sin\pi z$ are representative of the solutions
obtained with any other reasonable specific form of $\alpha(z)$.

At larger values of $|D|$, we see that there is a range of dynamo numbers, $-60 \ge D \ge -160$, over which the growth rate obtained numerically can be well approximated by a power law in $D$. Applying linear 
regression, we obtain the following approximation in this range:
\begin{equation}\label{g044}
\gamma \approx 0.44 \left|D\right|^{0.52}\,.
\end{equation}
Comparing this expression with equation~\eqref{eq:growthrate} we see that these are comparable if $\bar{\gamma_0} \approx 0.275$, which is similar to the value found by \citet{IRSF81}. The accuracy, and usefulness for applications, of the asymptotic solution for $|D|\gg1$ can be enhanced by replacing $D$ by $D-D\crit$ with $D\crit=-8$ in equation~\eqref{eq:growthrate}; the resulting dependence,
\begin{equation}\label{g0275}
\gamma \approx 0.275 \left[\alpha'(0)(D\crit-D)\right]^{1/2}\,,
\qquad
D\crit=-8,
\quad
-160\la D< D\crit\,,
\end{equation}
 is also shown in figure~\ref {fig:ln_gamma_vs_D}. 

The degree of agreement between the 
two
approximations is striking. Most notably, the growth rate predictions for $|D|\ll1$ and $|D|\gg1$ evidently remain reasonably accurate even far beyond their respective ranges of applicability. Such behaviour is not unusual in asymptotic solutions; in our case, it apparently results from the simple form of the eigenfunction (see below), as well as from the similarity of the scalings of $\gamma$ with $|D|^{1/2}$ in both asymptotic extremities.

\subsection{The eigenfunctions}\label{Eig}
\begin{figure}
\begin{center}
\includegraphics[width=0.4\textwidth,height=0.35\textwidth]{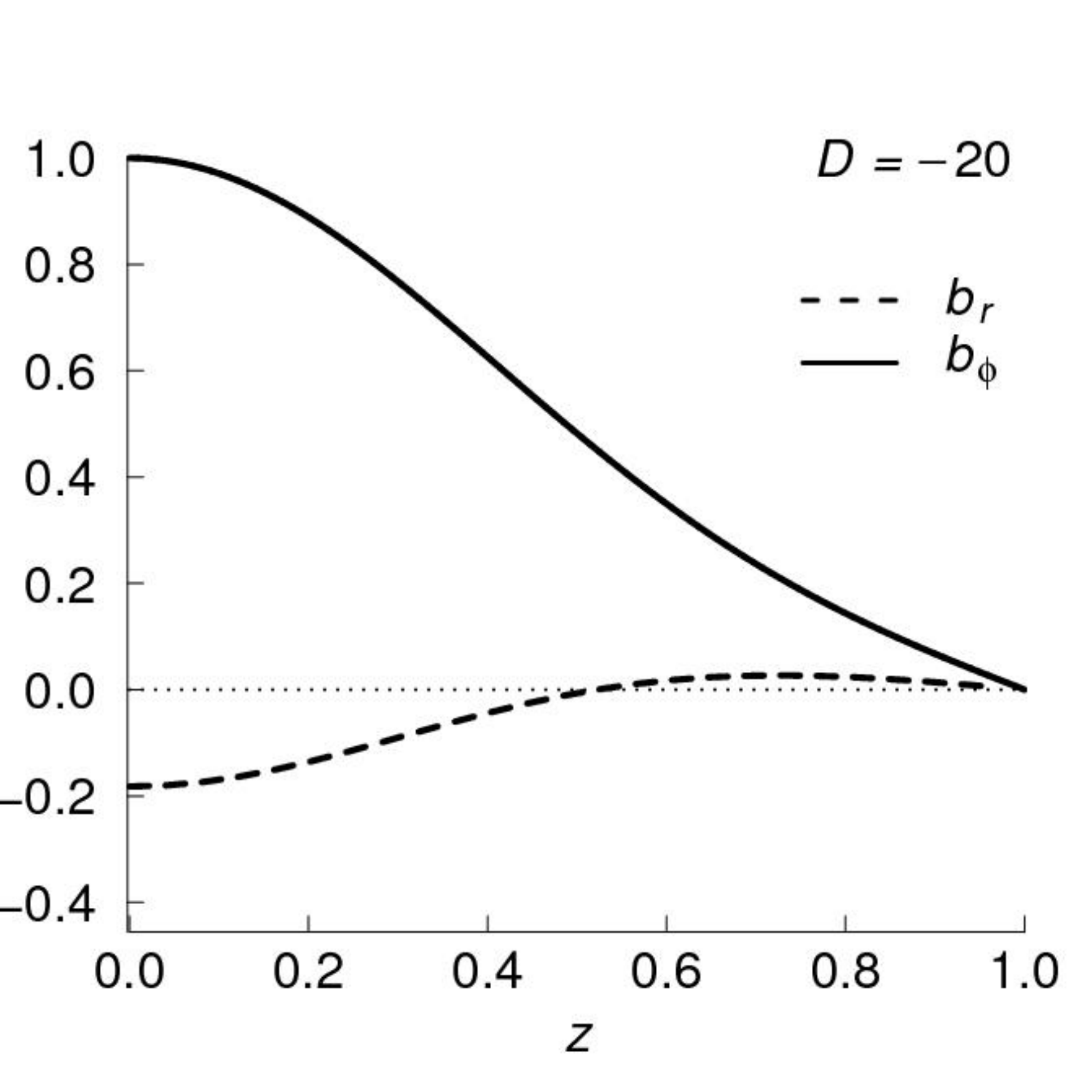}
\includegraphics[width=0.4\textwidth,height=0.35\textwidth]{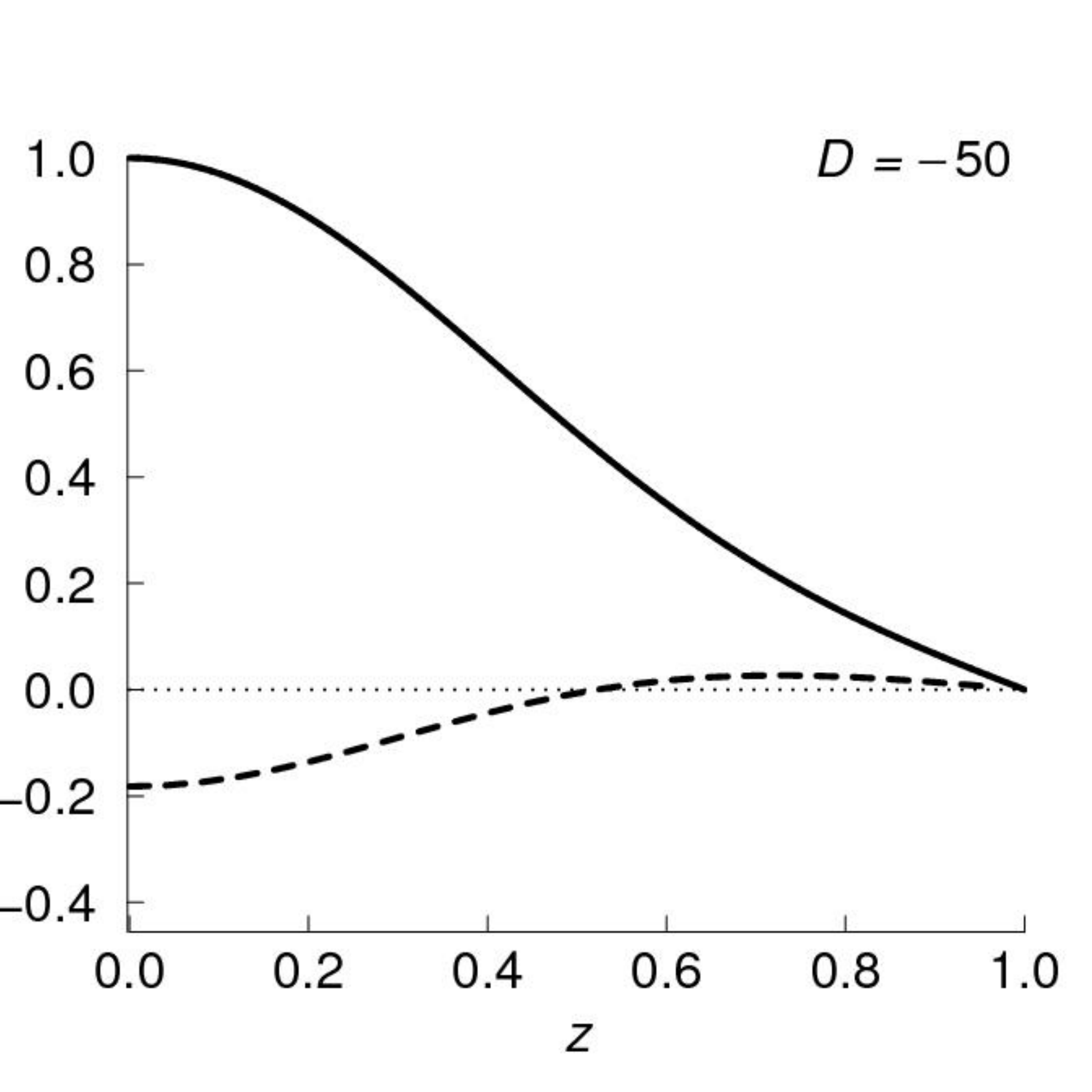}\\
\includegraphics[width=0.4\textwidth,height=0.35\textwidth]{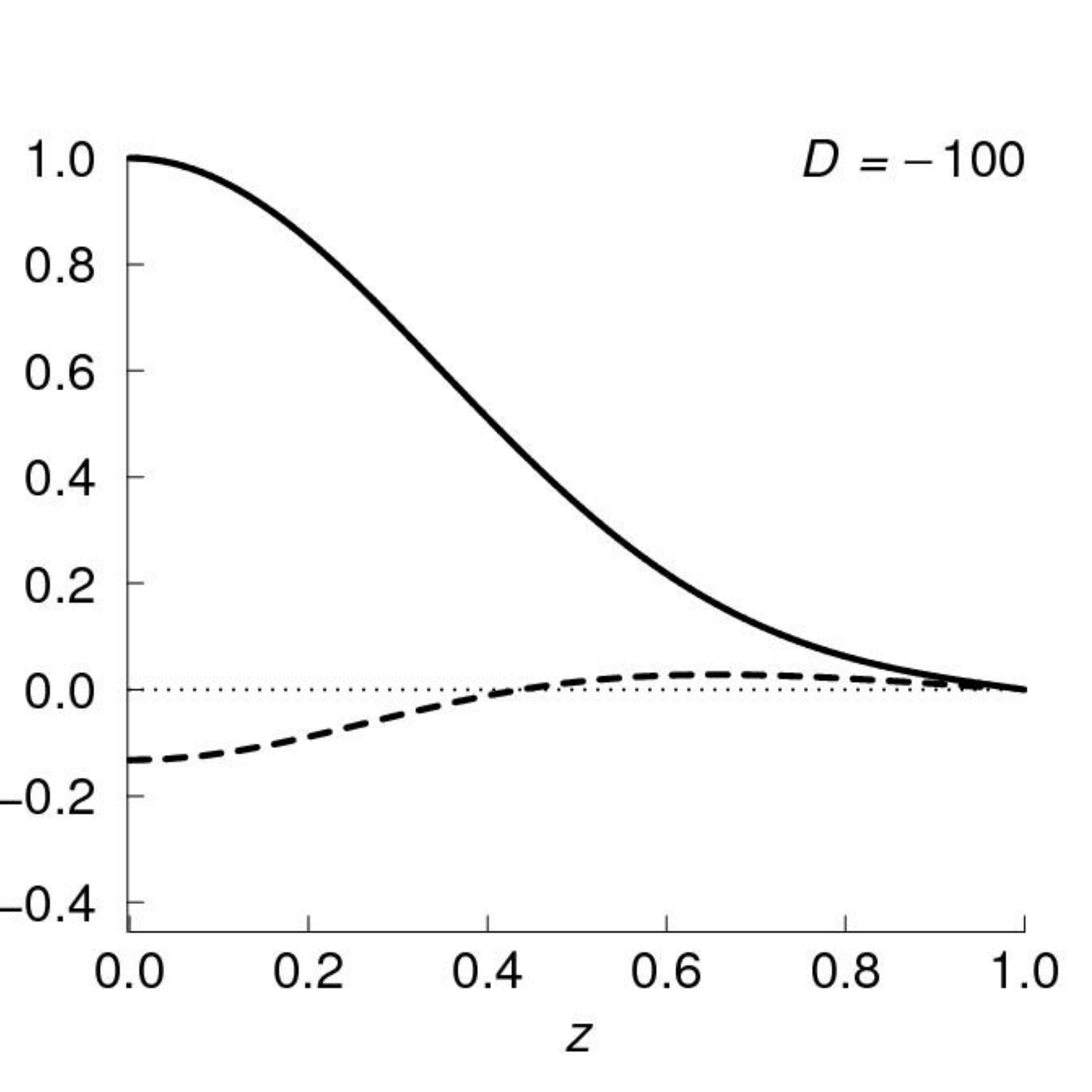}
\includegraphics[width=0.4\textwidth,height=0.35\textwidth]{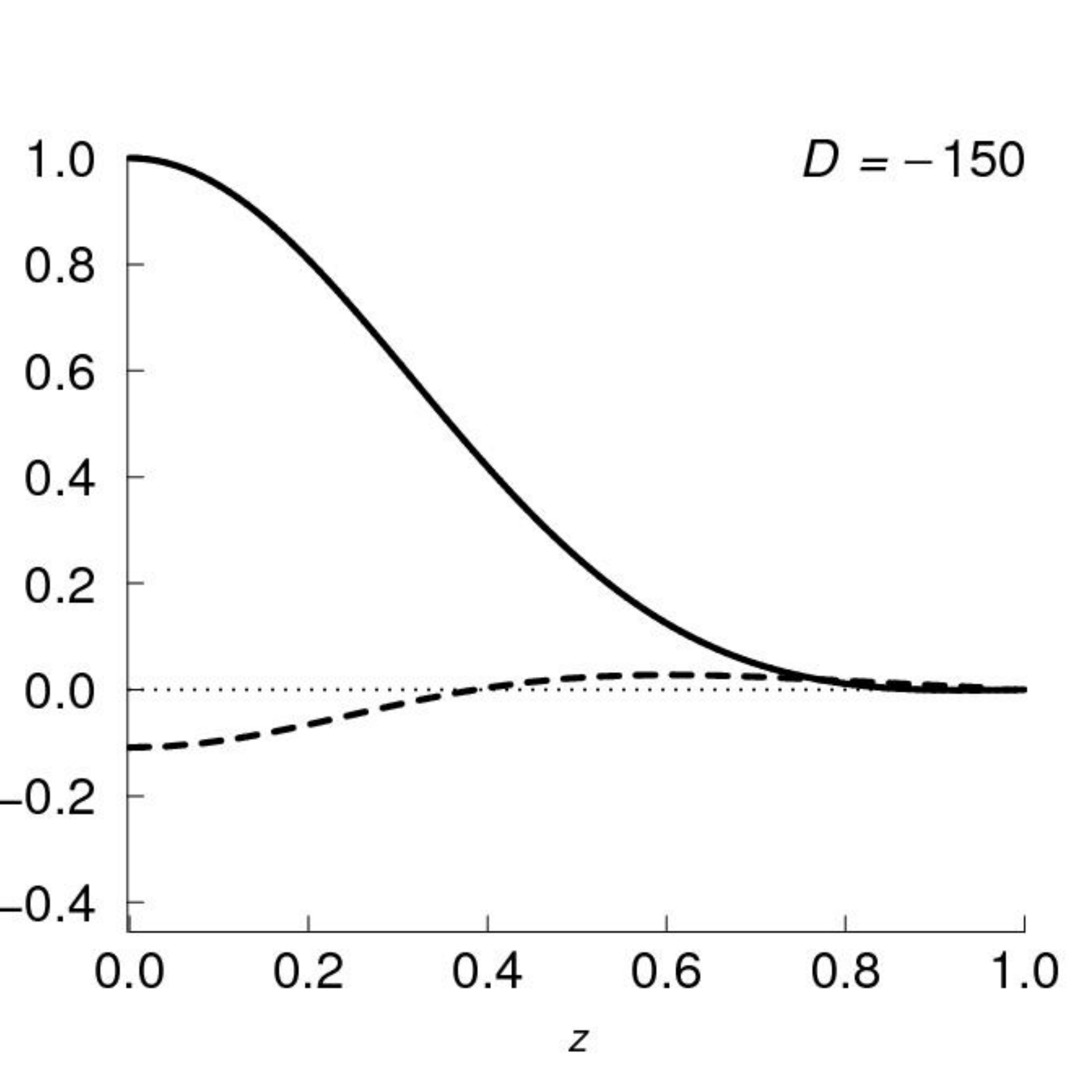}\\
\includegraphics[width=0.4\textwidth,height=0.35\textwidth]{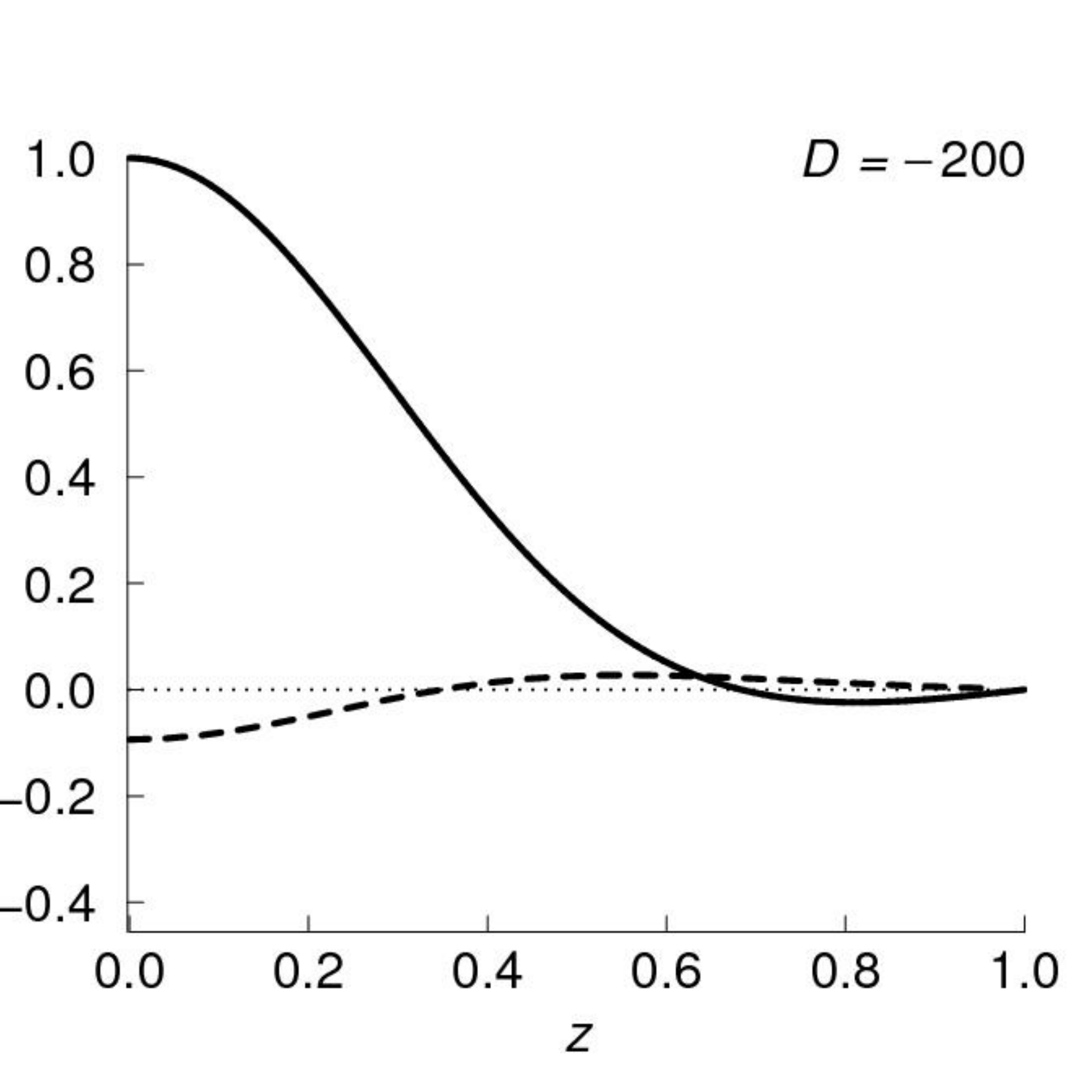}
\includegraphics[width=0.4\textwidth,height=0.35\textwidth]{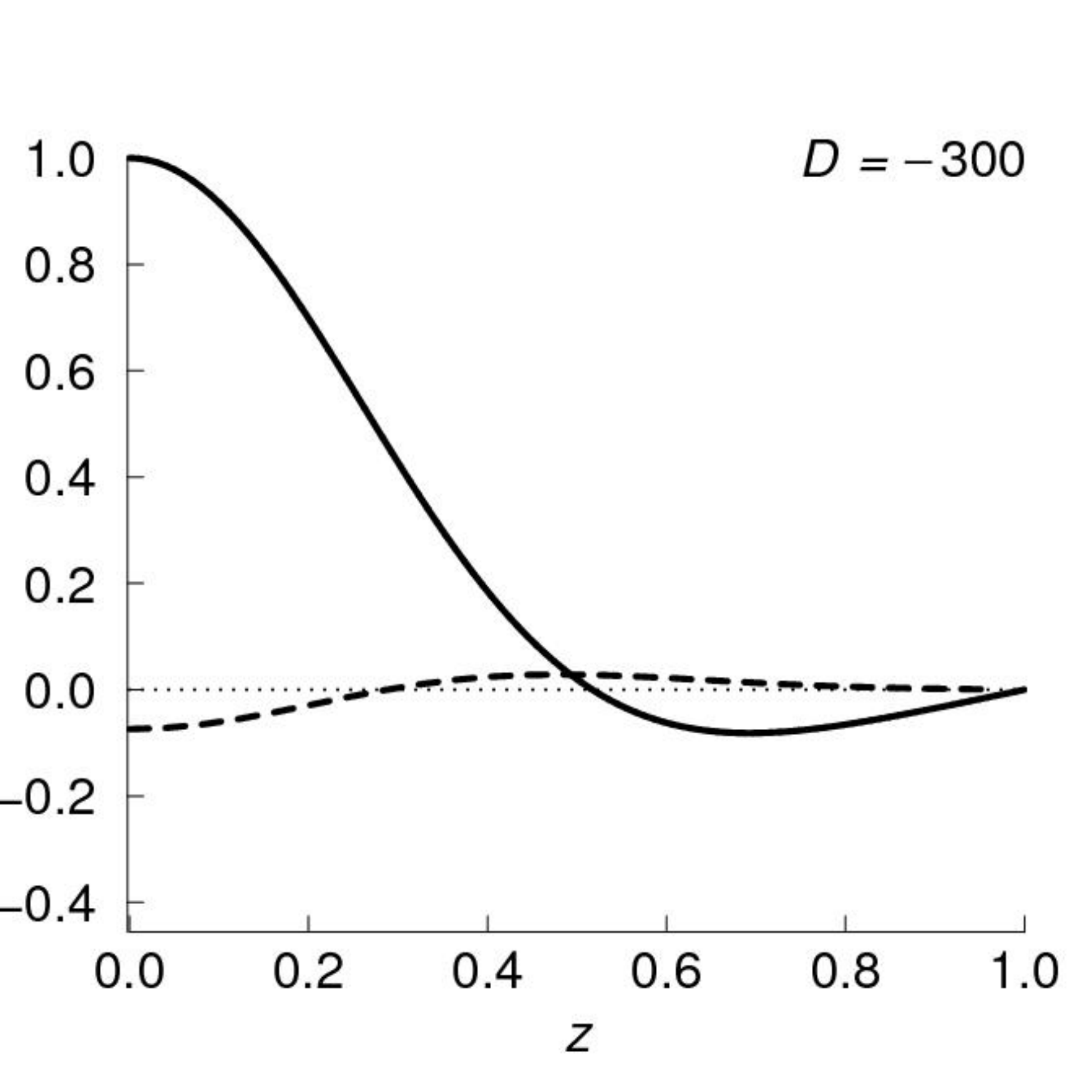}\\
\end{center}
\begin{center}
\caption{\label{fig:eigenfunctions1}The eigenfunctions $b_r$ 
(dashed) 
and $b_{\phi}$
(solid),
normalised to $b_{\phi}=1$ at $z=0$, for $D = -20,\ -50,\ -100,\ -150,\ -200$ and $-300$. 
Solutions become concentrated near the slab mid-plane, $z = 0$, as the dynamo number 
increases in magnitude. The radial 
magnetic 
field component $b_r$ always changes sign 
at a certain
value of $z$
in any growing solution, 
whereas $b_{\phi}$ only changes sign for large $\left|D\right|$. }
\end{center}
\end{figure}

Figure~\ref{fig:eigenfunctions1} shows the eigenfunctions obtained numerically 
for various values of the dynamo number. For ease of comparison, the eigenfunctions 
have all been normalised so that $b_{\phi}=1$ at $z=0$. 
For small values of $|D|$, $b_{\phi}$ decreases monotonically with increasing $z$. 
However, $b_r$ (which is negative at $z=0$) always changes sign at some point within 
the domain. The value of $z$ at which $b_r$ changes sign decreases with increasing $|D|$.
\citet{RTZS79} have clarified the importance of this feature for the dynamo action:
the change of sign of $b_r$ drives a diffusive magnetic flux from the dynamo region,
thus allowing magnetic field to grow. For larger values of $|D|$, there is also a sign change 
in $b_{\phi}$, although this always occurs at a larger value of $z$ than the zero of 
$b_r$. Notably, the non-monotonic behaviour of $b_\phi$ becomes pronounced for $D\leq -150$ when
the dependence of $\gamma$ on $|D|$ deviates from a power law. This signifies a progressive
reduction in the accuracy of the asymptotic solution, \eqref{g044} or \eqref{g0275}, 
as $|D|$ increases further. 

\begin{figure}
\begin{center}
\includegraphics[width=0.4\textwidth,height=0.35\textwidth]{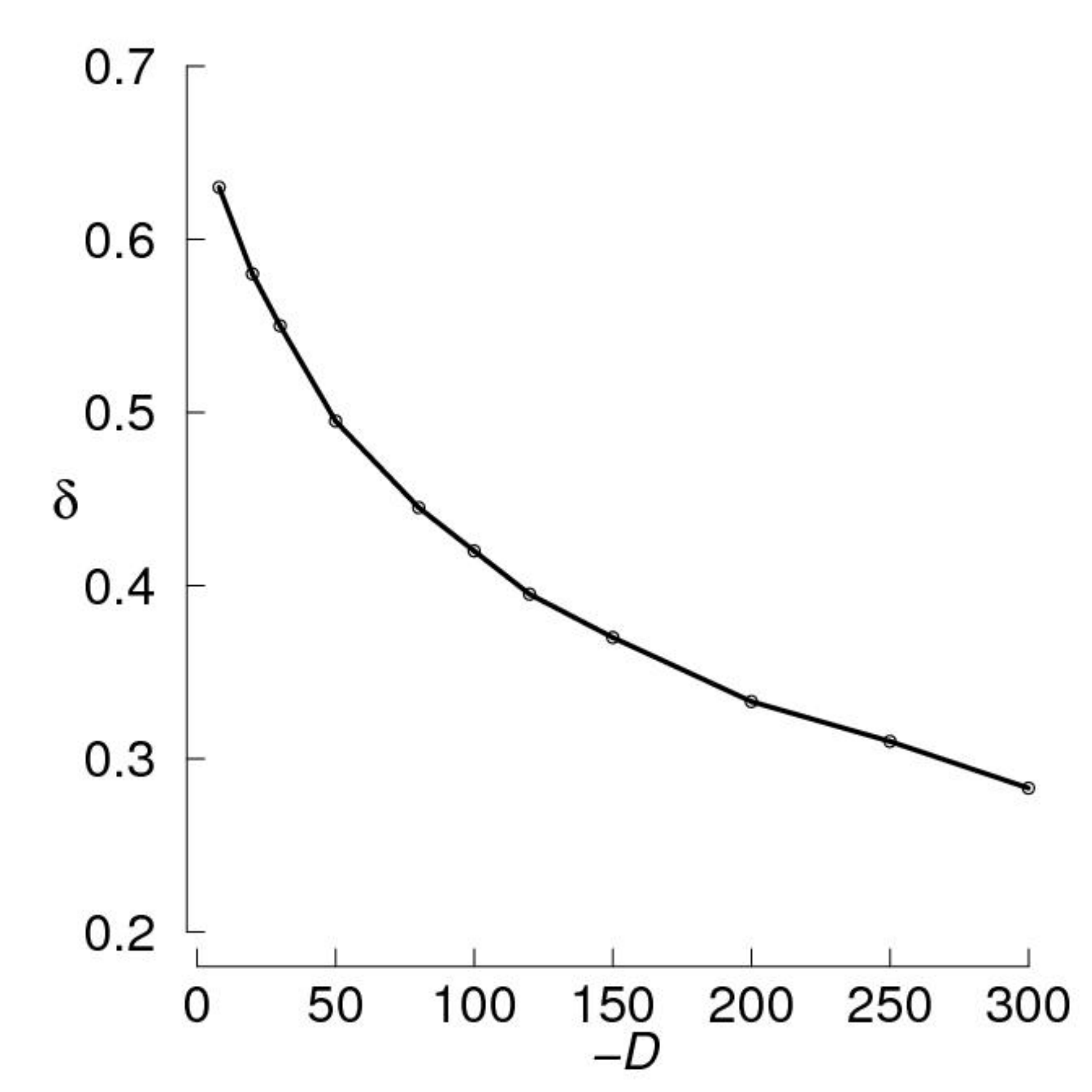}
\includegraphics[width=0.4\textwidth,height=0.35\textwidth]{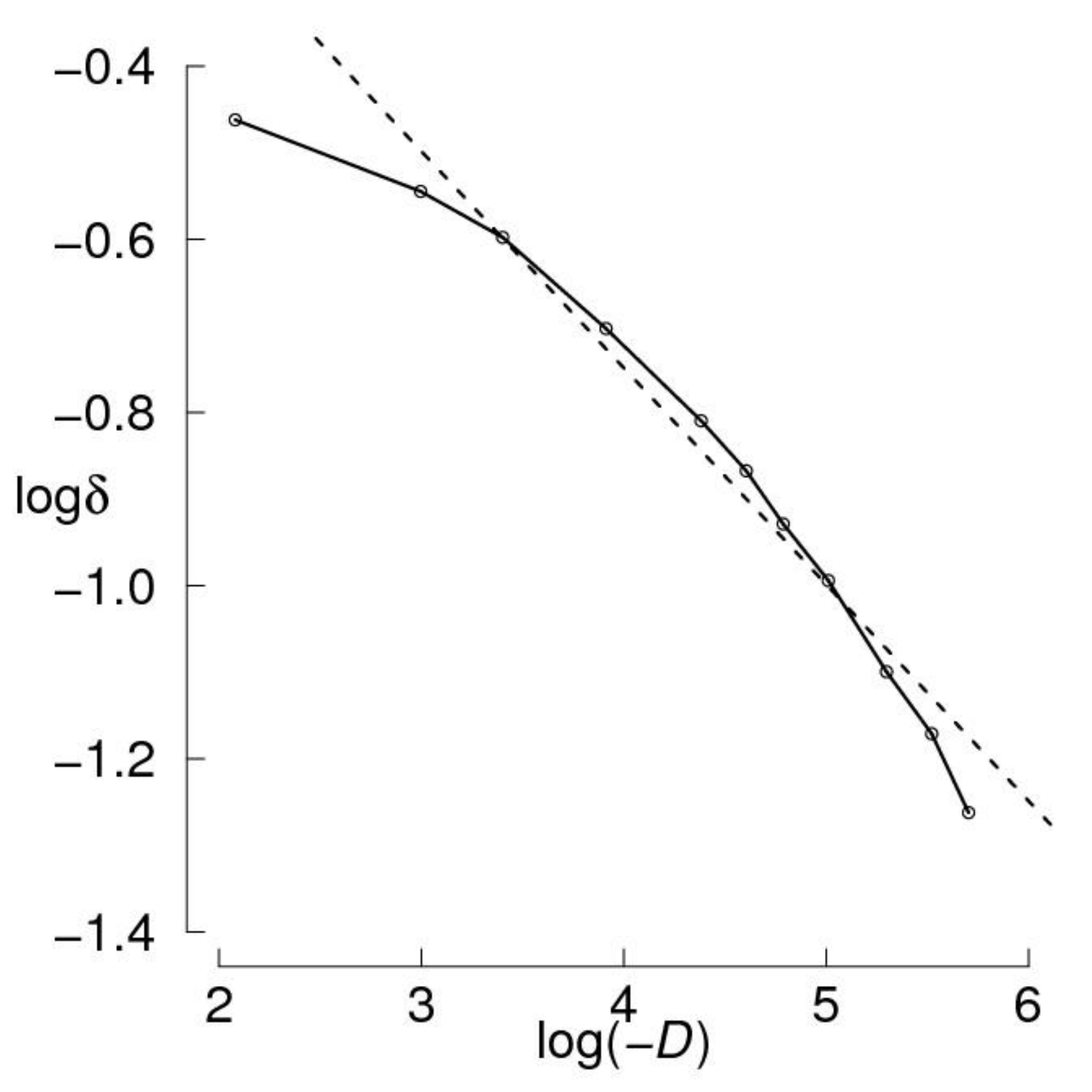}\end{center}
\begin{center}
\caption{\label{fig:delta_vs_D}
Left panel: The  half-width at half-maximum of the  eigenfunction, $\delta$, as a function of $|D|$. 
Right panel: The corresponding linear regression of $\ln \delta$ versus $\ln \left|D\right|$, 
$\ln \delta \approx -0.2387 \ln|D| +0.2079$ for $-160\leq D\leq-60$,
with the asymptotic result shown dashed}.
\end{center}
\end{figure}

Conforming to the concept of a boundary layer at $|D|\gg1$, 
the eigenfunction becomes more concentrated near 
the slab mid-plane as $|D|$ increases, so the scale in $z$, on which these 
functions vary, becomes smaller at larger dynamo numbers. 
This can be quantified in terms of $\delta$, the half-width at half-maximum of $b_{\phi}$:
\[
b_{\phi}(\delta)=\tfrac12 b_{\phi}(0)=\tfrac12\,.
\]
Figure~\ref{fig:delta_vs_D} shows the $D$-dependence of $\delta$ obtained numerically. The 
linear-regression fit of a power law has the form 
\[
\delta \approx 1.2\,\left|D\right|^{-0.24}\,,
\quad
-160\leq D <-60\,.
\]
This scaling is consistent with the asymptotic analysis, which suggests that the scale of 
the eigenfunction varies as $|D|^{-1/4}$.

\begin{figure}
\begin{center}
\includegraphics[width=0.4\textwidth,height=0.35\textwidth]{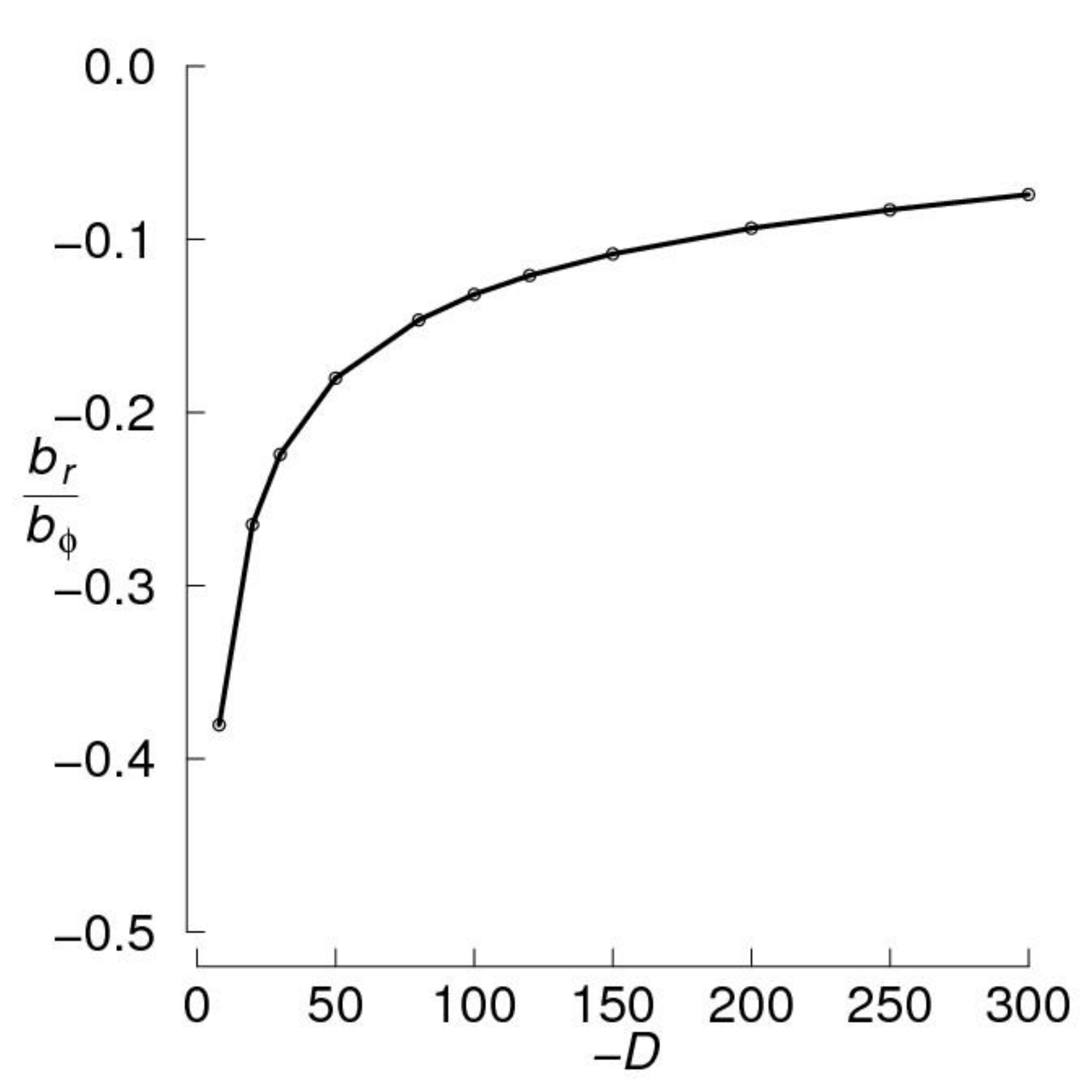}
\includegraphics[width=0.4\textwidth,height=0.35\textwidth]{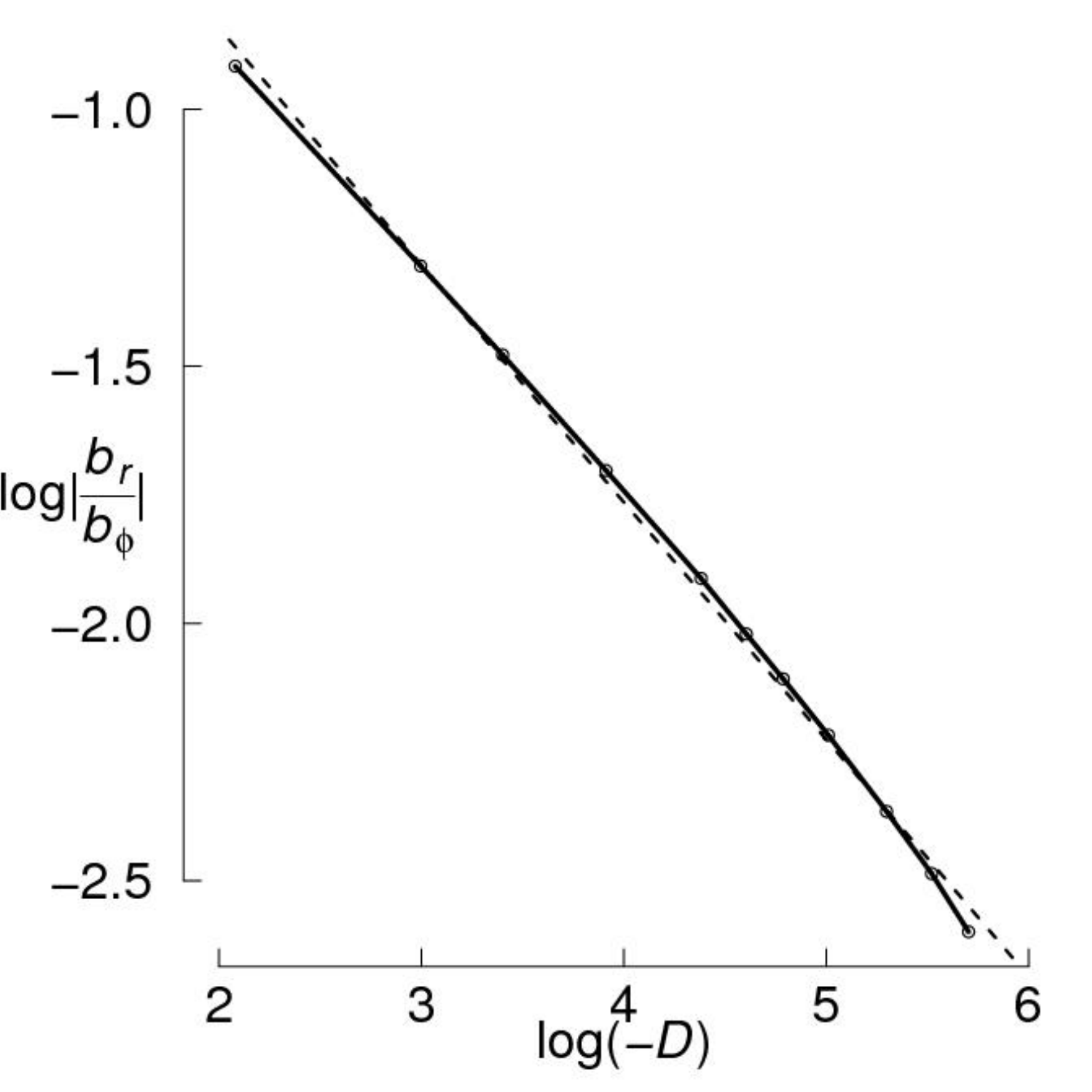}
\caption{\label{fig:pitch_angle}Left panel: The ratio $b_r/b_\phi$ at $z=0$, as 
a function of $-D$ for $D<0$. 
Right panel: The approximation obtained by linear regression, 
$\ln |b_r/b_\phi| \approx -0.461 \ln \left|D\right| + 0.0797$ at $z=0$,
with the asymptotic result shown dashed}.
\end{center}
\end{figure}

The magnetic pitch angle $p_B$, defined as $p_B = \arctan | b_r/b_\phi |$, is a readily 
observable quantity in galaxies, and it has been shown to be a sensitive diagnostic 
of the galactic dynamo action \citep[e.g.,][]{S07}. Figure~\ref{fig:pitch_angle} shows 
the related quantity, $ b_r/b_\phi$ evaluated at $z=0$. Its dependence on $D$
is very well approximated by a power law, 
\[
\frac{b_r}{b_\phi}\approx -1.1\, |D|^{-0.46}\,,
\quad
-160\leq D <-60\,.
\]
Again, the numerical results agree well with the asymptotic scaling 
$b_r/b_\phi\propto|D|^{-1/2}$. 

\begin{figure}
\begin{center}
\includegraphics[width=0.4\textwidth,height=0.35\textwidth]{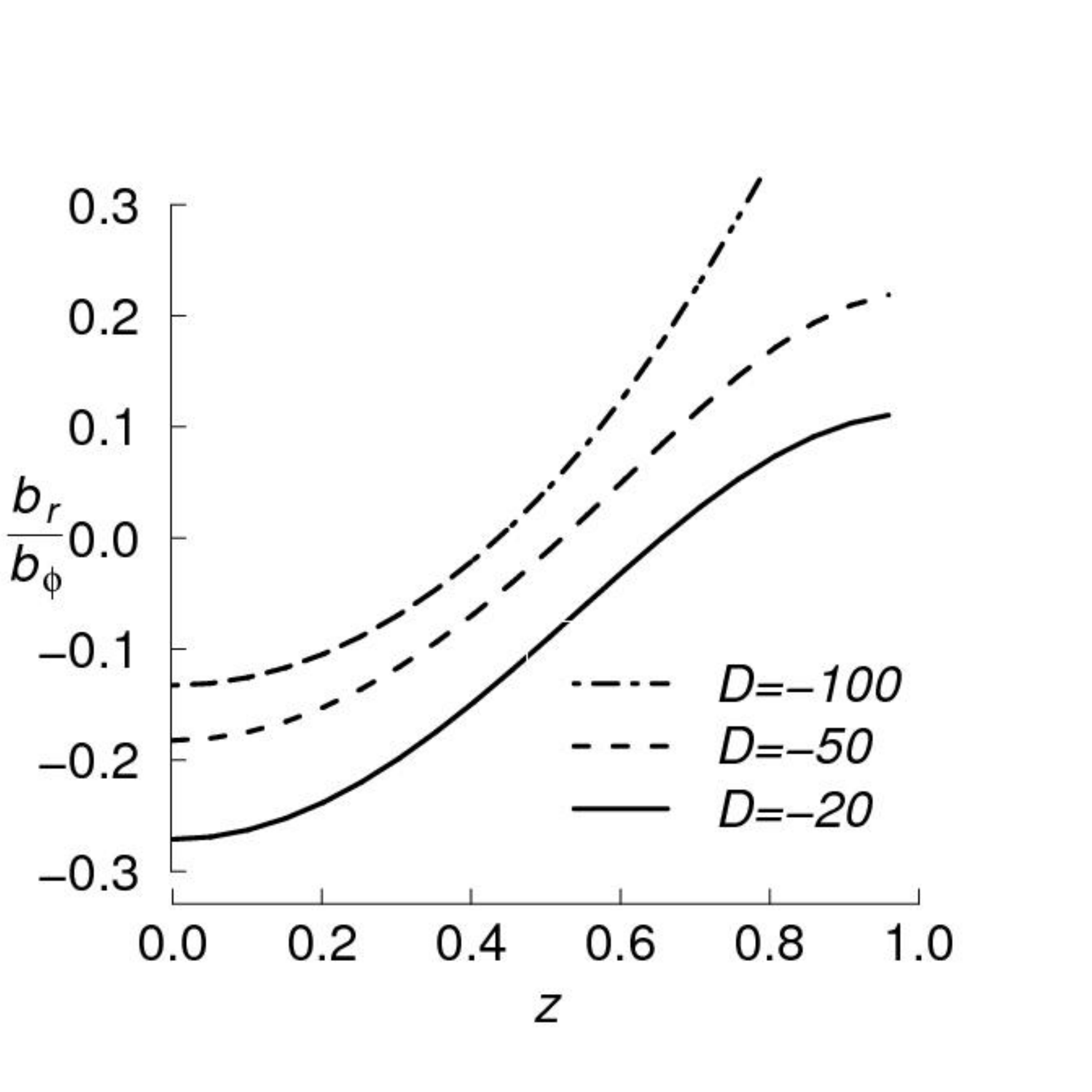}
\includegraphics[width=0.4\textwidth,height=0.35\textwidth]{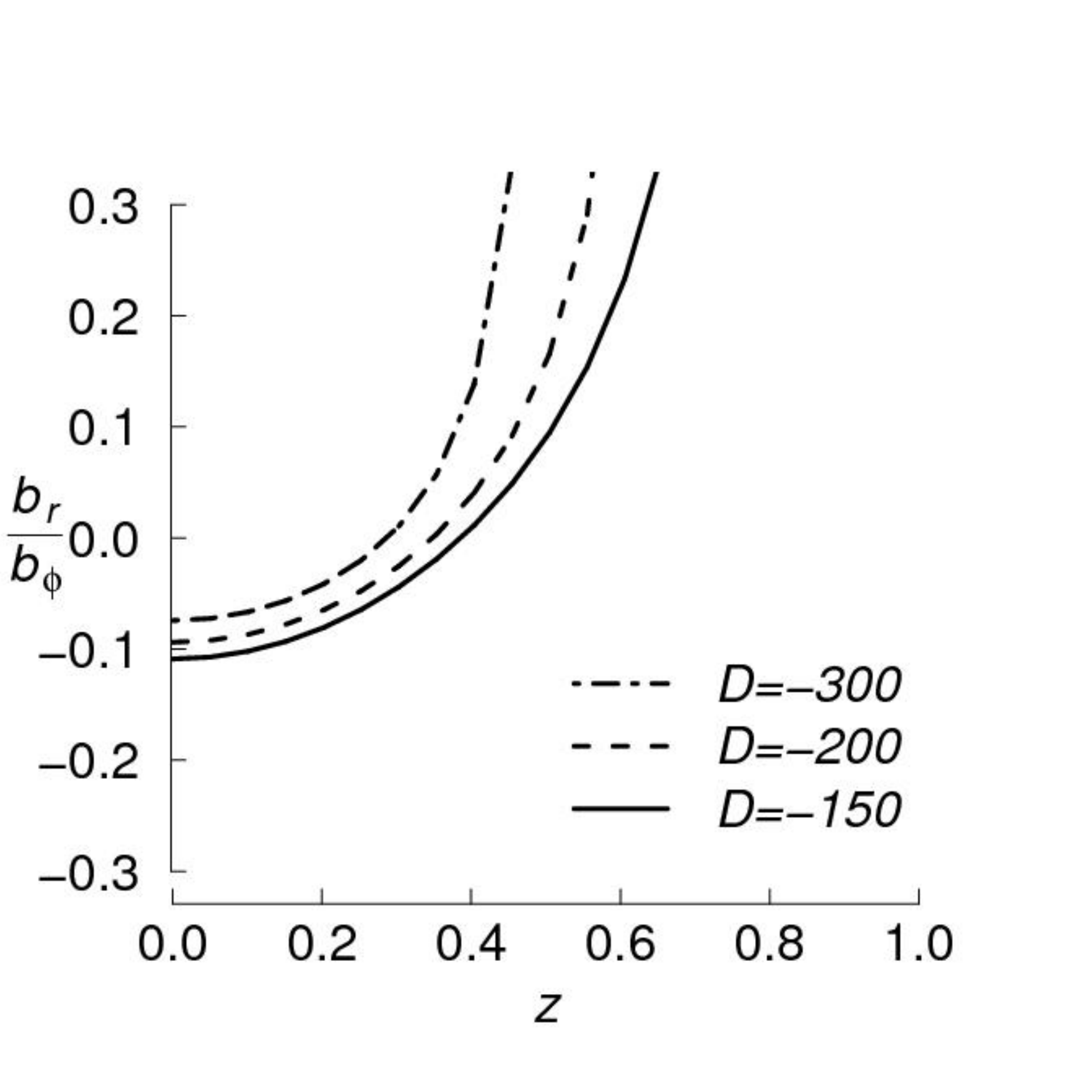}
\end{center}
\begin{center}
\caption{\label{BL} The ratio $b_r/b_\phi$ as a function of $z$
for $D=-20$,$-50$ and $-100$ (left-hand panel) and $D=-150$, $-200$ and $-300$
(right-hand panel), with line styles indicated in the figure legends.}
\end{center}
\end{figure}

In their analytic study, \citet{IRSF81} 
assumed that the eigenfunctions of $b_{\phi}$ and $b_r$ differed only by a constant of 
proportionality. This assumption allowed these authors to reduce the problem to a single
ordinary differential equation. To clarify whether such an assumption is justifiable,
we show in figure~\ref{BL} the ratio $b_r/b_\phi$ as a function of $z$. For the assumption 
$b_r/b_\phi=\mbox{const}$ to be valid, the relative change in this ratio should not
exceed $\mathrm{O}(|D|^{-1/4})$ within the boundary layer. The relative change in $b_r/b_\phi$ is 0.46, 
0.48 and 0.40 at $D=-50,-150$ and $-300$ even within a single length scale, 
$0\leq z \leq|D|^{-1/4}$, with $|D|^{-1/4}=0.38, 0.29$ and $0.24$, respectively. 
Thus, this assumption is dangerous at least, if applicable at all.

\begin{figure}
\begin{center}
\includegraphics[width=0.4\textwidth,height=0.35\textwidth]{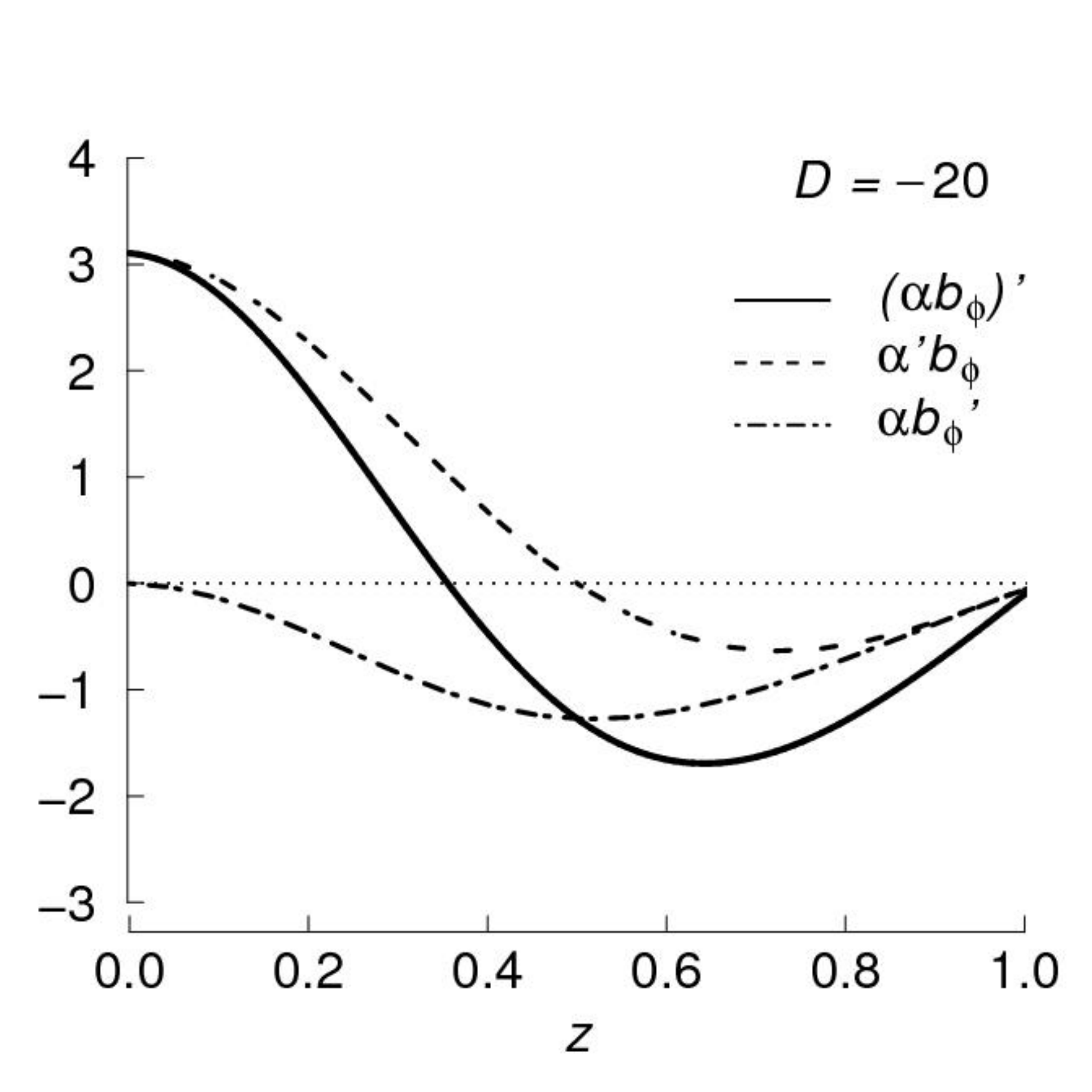}
\includegraphics[width=0.4\textwidth,height=0.35\textwidth]{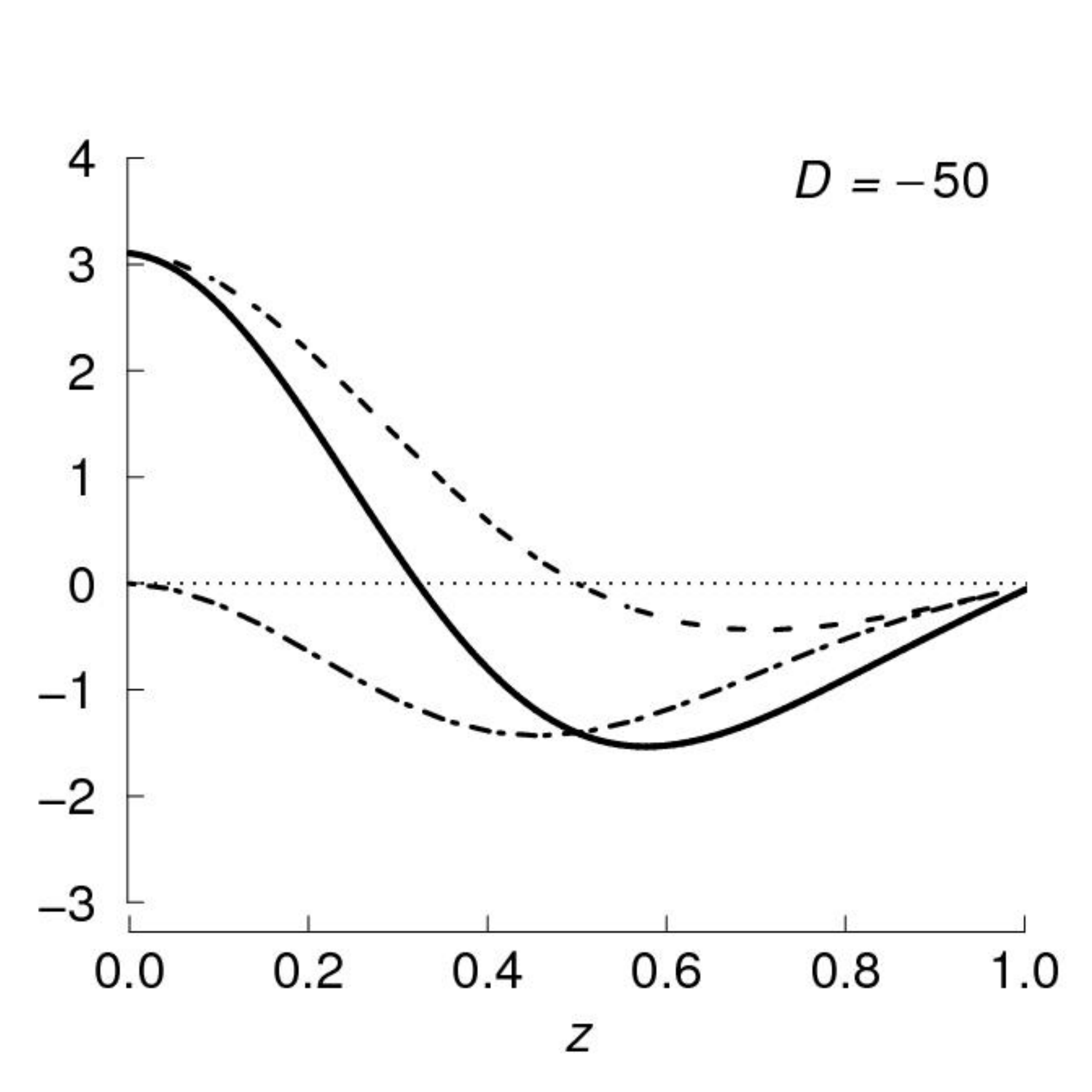}\\
\includegraphics[width=0.4\textwidth,height=0.35\textwidth]{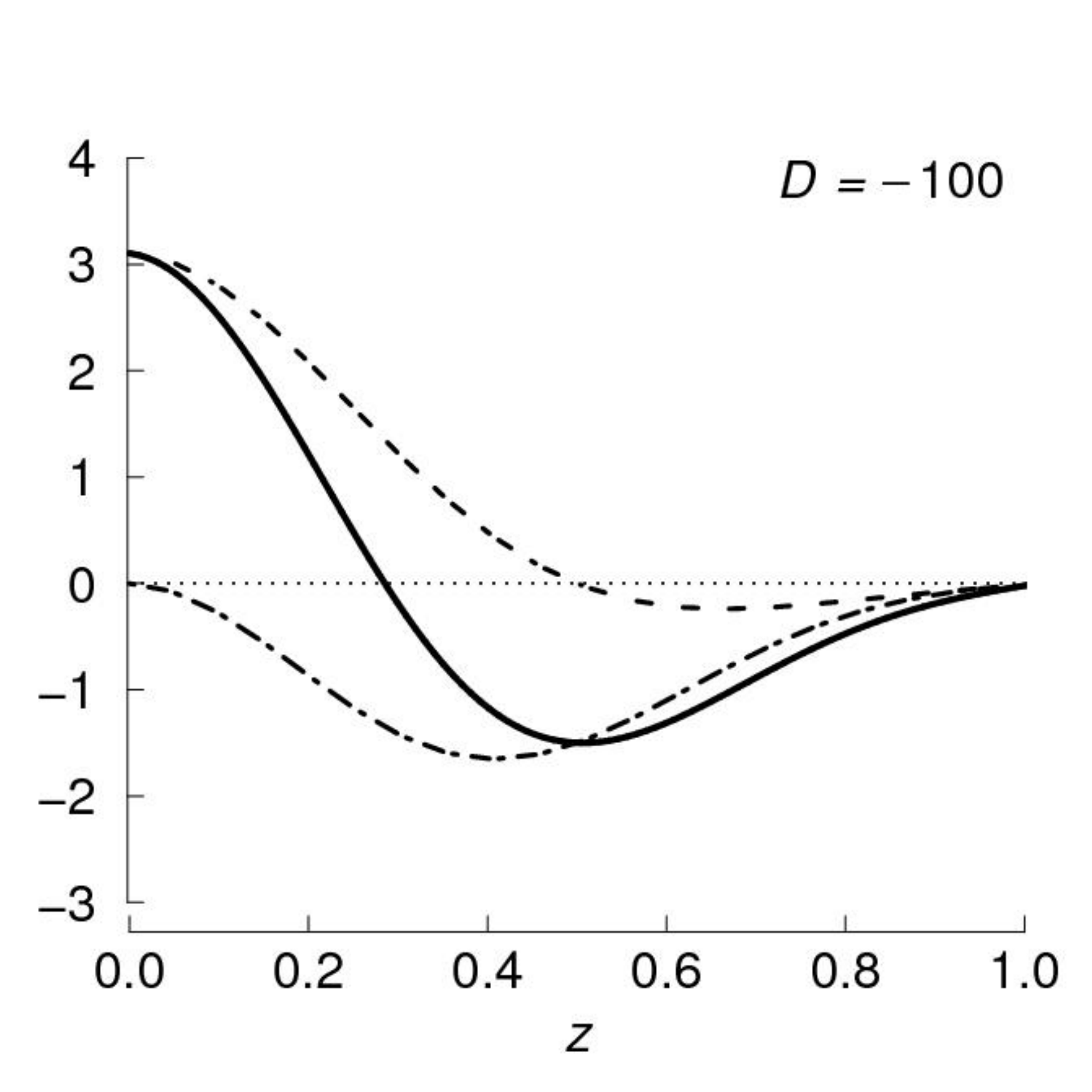}
\includegraphics[width=0.4\textwidth,height=0.35\textwidth]{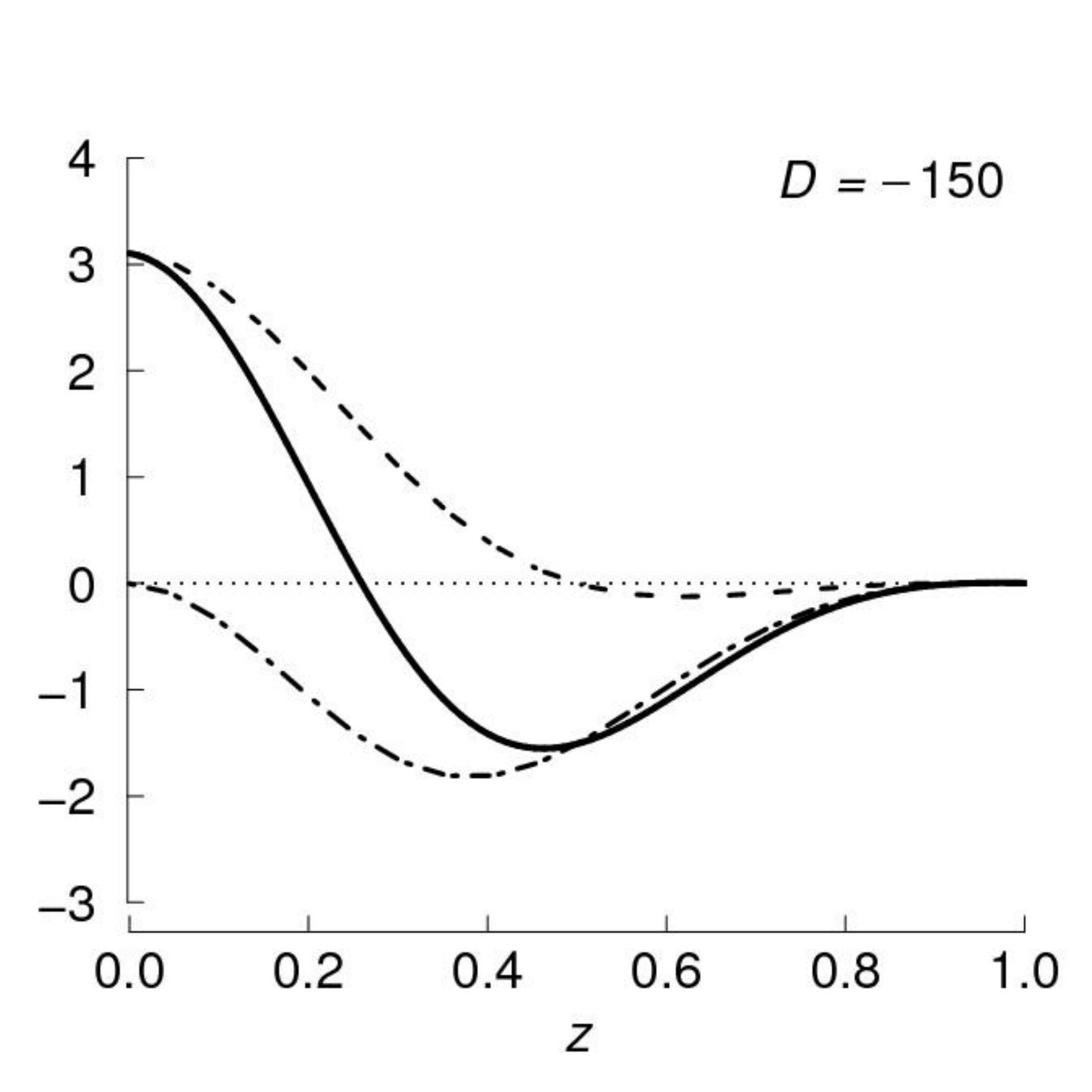}\\
\includegraphics[width=0.4\textwidth,height=0.35\textwidth]{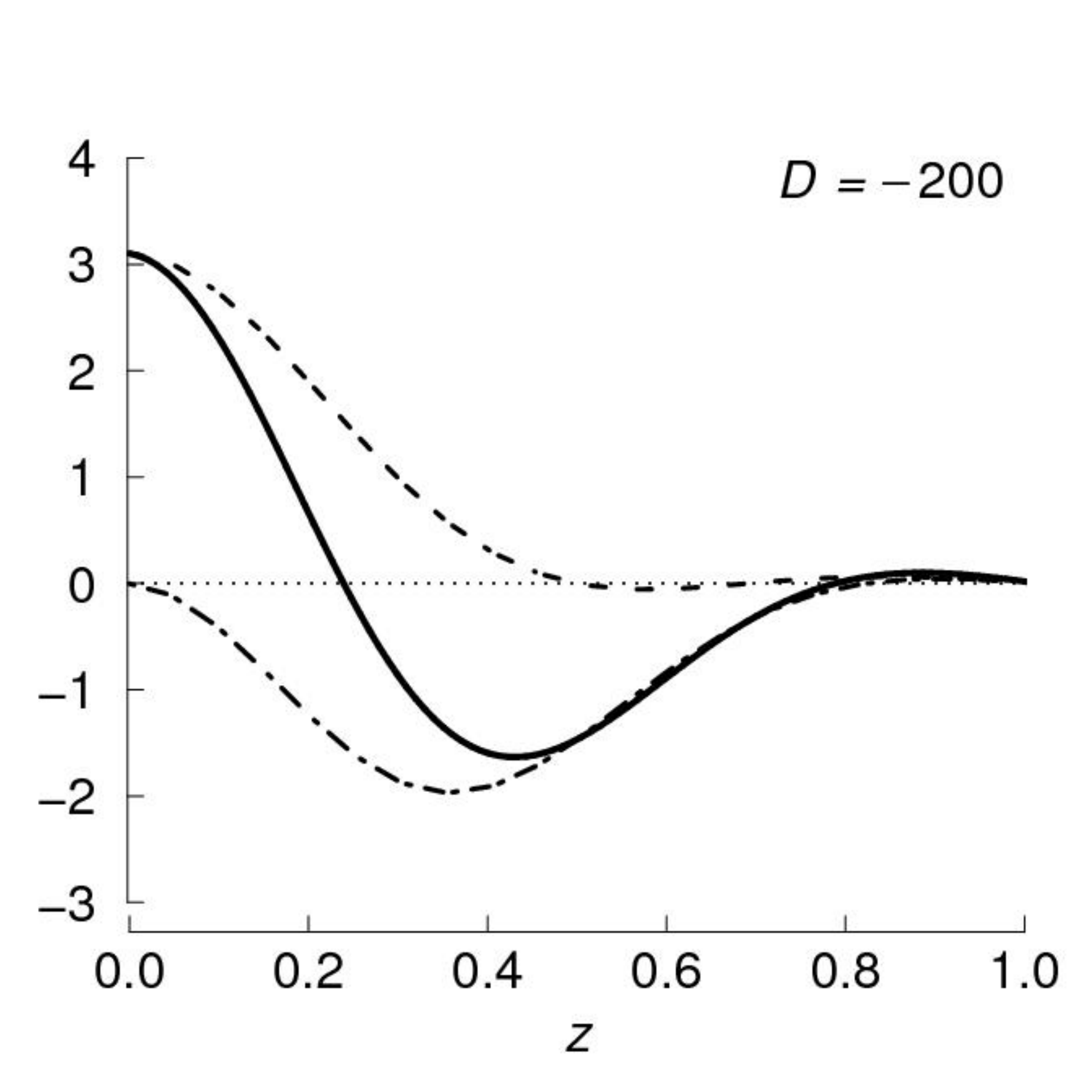}
\includegraphics[width=0.4\textwidth,height=0.35\textwidth]{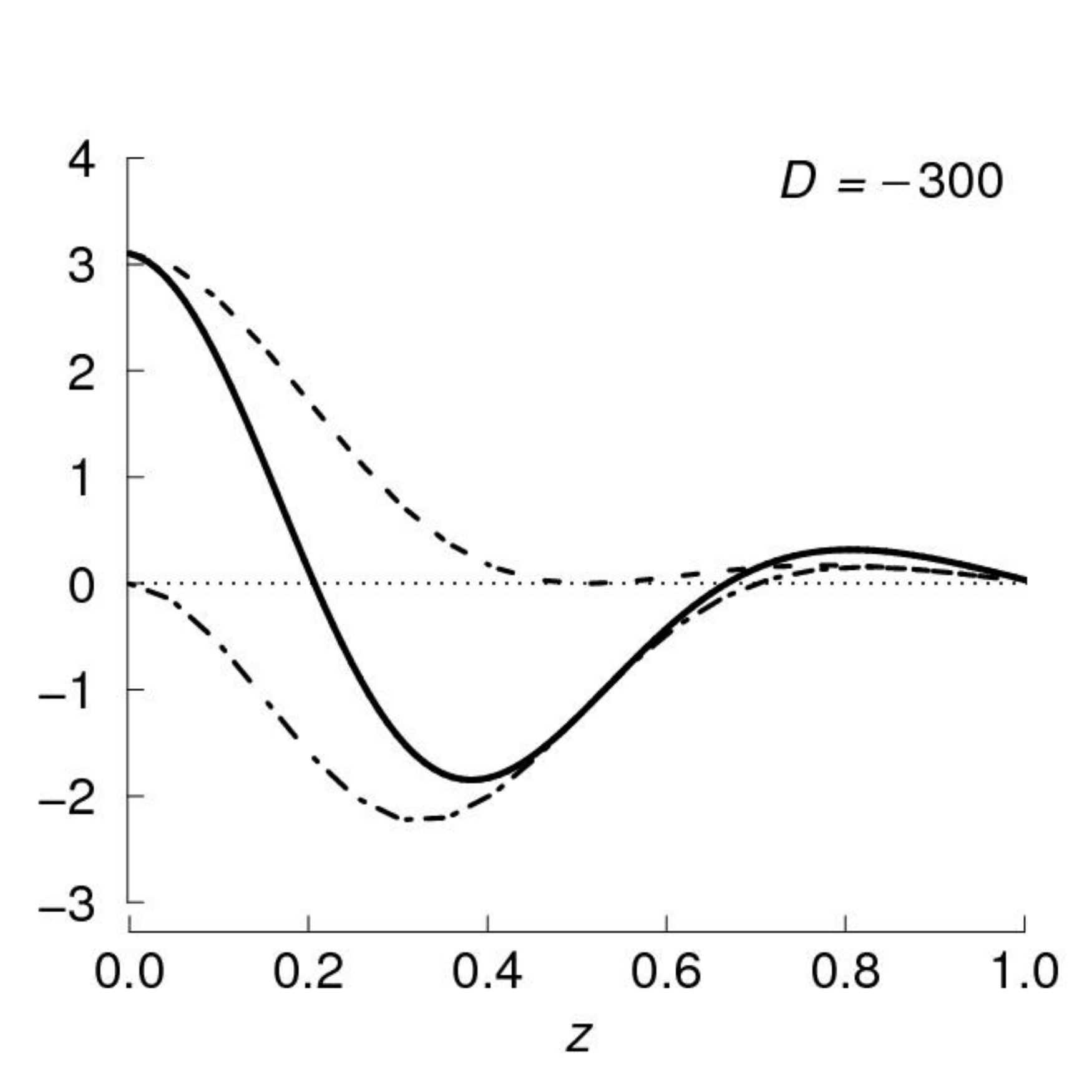}\\
\end{center}
\begin{center}
\caption{\label{fig:alphaB}The 
individual
terms representing the $\alpha$-effect,
$(\alpha b_\phi)'= \alpha' b_\phi+\alpha b_\phi'$, for $\alpha=\sin\pi z$ and
$D = -20,\ -50,\ -100,\ -150,\ -200$ and $-300$ (here prime denotes derivative in $z$). 
The term on the left-hand side
of the equation, $(\alpha b_\phi)'$, 
is shown solid, $\alpha' b_\phi$, dashed,
and $\alpha b_\phi'$,} dash-dotted. In each case, the curves have been normalised so that 
$(\alpha b_\phi)'=\pi$ at $z=0$. 
The contribution from $\alpha'b_{\phi}$ dominates near $z=0$ whereas 
$\alpha b_{\phi}'$ dominates at mid-depths. 

\end{center}
\end{figure}

\subsection{The inner working of the dynamo}\label{IW}   
Finally, we consider the role of the various contributions to the $\alpha$-effect in the dynamo 
mechanism. One of the main conclusions of the asymptotic analysis was that the two different parts of the 
$\alpha$-effect on the right-hand side of equation~(\ref{eq:alphab}) should (in general) be of a 
comparable magnitude in $|D|$, which means that neither should be neglected. 
Figure~\ref{fig:alphaB} shows the spatial dependence of the terms that represent the 
$\alpha$-effect in the governing equations. 
Since $\alpha(0)=0$ and $b'_\phi(0)=0$ , it is unsurprising that 
$(\alpha b_{\phi})' \approx \alpha' b_{\phi}$ in the 
immediate vicinity of $z=0$ (with prime denoting the derivative with respect to $z$). 
However, $|\alpha' b_{\phi}|$ decreases with $z$, 
with $\alpha b_{\phi}'$ making the dominant (negative) contribution to the $\alpha$-effect at 
larger $z$. This transition occurs at progressively smaller values of $z$ as $|D|$
increases, $(\alpha b_\phi)'=0$ at $z\approx0.26$ for $D=-150$ and $z\approx0.2$ for $D=-300$.  
With the spatial scale of the solution of order $|D|^{-1/4}\approx0.29$ for $D=-150$ and $0.24$
at $D=-300$, $(\alpha b_\phi)'$ changes sign within the boundary layer. Clearly the neglect of 
either of these terms would affect the solution.

\begin{figure}
\begin{center}
\includegraphics[width=0.25\textwidth]{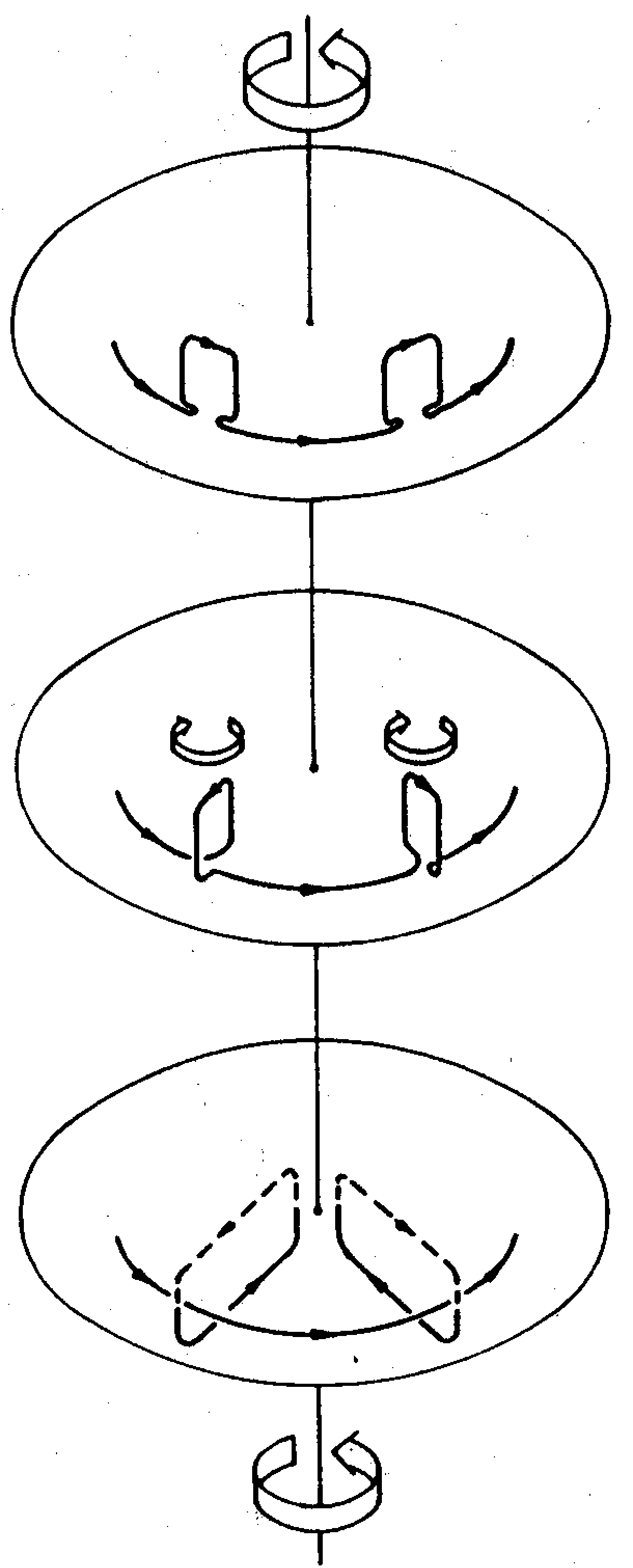}
\end{center}
\begin{center}
\caption{\label{cartoon}
The action of the $\alpha$-effect on the azimuthal magnetic field
produces a poloidal field inside the disc by twisting turbulent magnetic
loops in a systematic manner. The upper part of the meridional magnetic line
(shown dashed in the lowest panel) leaves the disc due to the diffusive flux. 
Only the upper half-space is shown in all the panels; 
mirror-antisymmetric events occur in the lower half-space
\citep[after][]{RSS88}.
}
\end{center}
\end{figure}

Since $b_r<0$, and $\de^2 b_r/\de z^2>0$ near $z=0$, it follows from equation~\eqref{eq:br} 
that $\gamma>0$ because $(\alpha b_\phi)'>0$ near $z=0$. With $\de b_r/\de z>0$, 
its diffusive flux,$-\de b_r/\de z$, is directed towards the midplane. 
At larger values of $z$, both $(\alpha b_\phi)'$ and $\de^2 b_r/\de z^2$ change sign,
the former because $\alpha b_\phi'$ becomes dominant (see figure~\ref{fig:alphaB}), and
the latter, due to a change in the curvature of $b_r(z)$ (see
figure~\ref{fig:eigenfunctions1}). Now, $\de b_r/\de z<0$
and the diffusive flux of $b_r$ is directed out of the dynamo-active region. 
\citet{RTZS79} 
discuss the importance of the outward flux through the surface (and the change of sign of $b_r$
within the slab) for the dynamo action: magnetic field produced by the $\alpha$-effect near the surface
of the slab (shown dashed in the lowest panel of figure~\ref{cartoon}) must leave through 
its surface to ensure that the oppositely directed net poloidal magnetic field
(shown solid) can grow within the slab.

Thus, the term $\alpha b_\phi'$ opposes the dynamo action by reducing the rate of growth of the radial magnetic field near the mid-plane of the slab. Its role in the dynamo action consists of driving the outward diffusive flux of magnetic field by changing the sign of $b_r$ closer to the slab surface. It is clear that such a surface-directed flux must occur within the boundary layer to allow magnetic field to grow, and hence this term cannot be neglected even within the boundary layer. We show in Appendix~\ref{SC} that the dynamo equations have only trivial solutions for $D<0$ if this term is neglected.

However, driving the flux to the surface comes at a price for the dynamo action.
For $D\la-160$, the negative contribution of $\alpha b_\phi'$ becomes so strong
that the dynamo action is affected and $\gamma$ increases with $|D|$ more gradually, and eventually
even decreases for $D<-205$, as shown in figure~\ref{fig:ln_gamma_vs_D}. Other modes, which become
dominant at still larger values of $|D|$, require a different balance of terms and can
prosper.


\section{Conclusions and discussion}\label{Disc}
This study has shown that the asymptotic and perturbation solutions of the mean-field $\alpha\omega$-dynamo 
are remarkably accurate in a wide range of dynamo numbers that covers those 
encountered in spiral galaxies and accretion discs near black holes and protostars. 
This provides a firm basis for the exploration of the origin of large-scale magnetic fields in those objects. We have also clarified the role of various terms in the dynamo equations, especially
$\alpha\,\de B_\phi/\de z$ and $B_\phi\,\de\alpha/\de z$, and demonstrated that neither 
of these terms should be neglected in a boundary-layer analysis. Indeed, as shown in Appendix~\ref{SC}, the neglect of the $\alpha\,\de B_\phi/\de z$ term implies that the boundary-layer equations only have trivial solutions for $D<0$. As emphasized by \citet{S95}, the solutions that are considered here are intermediate asymptotics
as they apply at $D\ll-1$ but only as long as the eigenvalue $\gamma$ remains real
and the quadrupolar mode is strongly dominant. We would therefore expect this analysis to break down when $-D$ exceeds $300$--500, although it should be stressed again that these very large dynamo numbers correspond to a less astrophysically-relevant parameter regime. Having explored the details of the dynamo process, we have also clarified the specific reasons for the deviation from the quadrupolar boundary-layer asymptotics as $|D|$ increases. As a result of this study, we would argue that the asymptotic behaviour of the $\alpha\omega$-dynamo in a slab for $D\ll-1$ is now clear and free of controversy.
 
When considering boundary-layer asymptotics for an $\alpha\omega$-dynamo model, it is important to bear in mind some of the constraints associated with the issue of scale separation. The mean-field dynamo equations are usually derived by assuming that the turbulent scale $l$ is much smaller than the spatial scale of the mean magnetic field (which is $h|D|^{-1/4}$ in this boundary-layer analysis). This simple argument implies that $|D|\ll(h/l)^4$ is a necessary condition for scale separation in this case. In spiral galaxies, we have $h=0.5\kpc$ and $l=0.1\kpc$ \citep{RSS88,BBMSS96}, which implies that this constraint corresponds to $|D|\ll600$. In the range of the values for $|D|$ for which quadrupolar solutions are preferred, this constraint is largely satisfied. However, because dipolar modes are only preferred at much larger values of $|D|$, this scale-separation constraint does become more significant in that case. This suggests that similar asymptotic methods may not be applicable to dipolar modes in a thin disc. Having said that it is worth noting that the identification of scale separation depends rather crucially upon the method of averaging that is adopted. Using spatial averages, it can be rather difficult to detect scale separation in numerical simulations \citep[e.g.,][]{BRS08,HP13}. On the other hand,  \citet{GSSFM13b} have shown that a Gaussian smoothing method (adapted to satisfy the Reynolds averaging rules) may be a more effective way of separating out the large-scale field from the small-scale fluctuations in turbulent flows. So the range of dynamo numbers for which such an asymptotic approach is applicable may be broader than that suggested by a simple two-scale analysis. 
 
It would be possible to extend this work by including nonlinearities into the governing equations, and this is an area that we intend to explore. Most natural dynamos are in a nonlinear, saturated state where the exponential growth of the magnetic field, characteristic of a kinematic dynamo, has been quenched by the action of the Lorentz force on the velocity field. For this reason, kinematic solutions are of limited interest in stellar and planetary dynamos, where the dynamo time scale is much shorter than the lifetime of the host object \citep{BKMMT89}. However, the situation is not so clear cut in the galactic context, where the $\ex$-folding time of the mean-field dynamo can be just 20--30 times shorter than the galactic lifetime \citep{RSS88}, so we would expect any nonlinear quenching mechanisms to operate over rather long timescales. In addition, large-scale magnetic fields in young galaxies have been observed, where the dynamo may still be in the kinematic stage \citep{BBMSS96}. In this context, kinematic dynamo models can play a very important role in explaining observations. So, although it will be interesting to investigate nonlinear effects in future work, our expectation is that kinematic models will also continue to provide new insights into galactic dynamo theory.

\section*{Acknowledgements}
We are grateful to Axel Brandenburg and Dmitry Sokoloff for useful comments.
YJ acknowledges financial support of the School of Mathematics and Statistics, Newcastle University,  
in the form of an undergraduate summer bursary. AS 
was supported by the Leverhulme Trust via Research Grant RPG-097.
PB and AS acknowledge financial support of the STFC via Research Grant ST/L005549/1.  

\appendices
\section{Simplified cases}\label{SC}
In this Appendix, we briefly revisit the boundary layer equations that were derived 
in section~\ref{ASDE}, focusing upon the effects of neglecting one of the terms on 
the right-hand side of Equation~(\ref{eq:alphab}). We should stress that a simplification 
of this kind is difficult to justify for a smoothly varying $\alpha$-effect. However, a 
number of idealised models have adopted functional forms for $\alpha(z)$ that cannot 
be represented by a Taylor series in the vicinity of $z=0$ \citep[see, e.g.,][]{M78,RTZS79}, 
in which case it may be appropriate to simplify the governing equations in this way.  

We first consider the consequences of the assumption 
$|b_\phi\, \upartial \alpha/\upartial z| \gg |\alpha\, \upartial b_\phi/\upartial z|$
made by \citet{IRSF81} and (in a slightly different context) \citet{S95}. 
Neglecting $\alpha\, \upartial b_\phi/\upartial z$ might appear justifiable
when $|\upartial\alpha/\upartial z|=\mathrm{O}(1)$ and $|\alpha|\ll1$ within the boundary layer.
In this case, equations~(\ref{eq:blf_1}) and~(\ref{eq:blg_1}) reduce to
\begin{align}
\Gamma_0 R &= - \Phi+ \frac{{\mathrm{d}^2} R}{\mathrm{d} s^2}\,  , \\
\Gamma_0 \Phi &= \mbox{sign}(D)\, R + \frac{{\mathrm{d}^2}\Phi }{\mathrm{d} s^2}\,,
\end{align}
with the boundary conditions \eqref{bsc}. In terms of the new variables,
\[
\Theta=R+\Phi\sqrt{-\mbox{sign}(D)}\,,
\qquad
\Psi=R-\Phi\sqrt{-\mbox{sign}(D)}\,,
\]
the equations of the system decouple:
\[
\Gamma_0\Theta=-\Theta[-\mbox{sign}(D)]^{3/2}+\frac{{\mathrm{d}^2}\Theta }{\mathrm{d} s^2}\,,
\] 
with the boundary conditions
\[
\frac{\mathrm{d}\Theta}{\mathrm{d} s}=0 \quad\mbox{at}\ s=0\,,
\qquad |\Theta|\to0\quad\mbox{for}\ s\to\infty\,,
\]
and likewise for $\Psi$. We note that $\Theta$ and $\Psi$ remain real for $D<0$ 
but become complex for $D>0$, with $\Psi=\Theta^*$ (asterisk denotes complex conjugate). 

Equations for $\Theta$ and $\Psi$ can easily be solved exactly, with the boundary condition as
$s\to\infty$ used to determine $\Gamma_0$. Thus, any further asymptotic analysis would be 
redundant \textit{if\/} this simplification was acceptable. Moreover, it can easily be seen that, for $D<0$, the only solution that satisfies the boundary condition at $s=0$ is $\Theta=A\cosh\left(s\sqrt{1+\Gamma_0}\right)$ with an arbitrary 
constant $A$. This implies that the boundary value problem with $|\Theta|\to0$ for $s\to\infty$ has only trivial
solutions. In many respects, this result is unsurprising. As we have shown in section~\ref{IW}, the term $\alpha\,\upartial b_\phi/\upartial z$ is essential for the dynamo action even if its role 
is more subtle than that of the other term representing the $\alpha$-effect.

If, on the other hand, it can be assumed that 
$|b_{\phi}\, \upartial \alpha/\upartial z| \ll |\alpha\, \upartial b_\phi/\upartial z|$, 
equations~(\ref{eq:blf_1}) and~(\ref{eq:blg_1}) become
\begin{align}
\Gamma_0 R &= - s\frac{\mathrm{d}\Phi}{\mathrm{d}s}+ \frac{\mathrm{d}^2R}{\mathrm{d}s^2}, \\
\Gamma_0 \Phi &= \mbox{sign}(D)\,R + \frac{\mathrm{d}^2\Phi}{\mathrm{d}s^2},
\end{align}
with the same boundary conditions as before. These equations are not much simpler than those 
of the full problem, so there are no clear benefits to neglecting this part of the 
$\alpha$-effect term either. 

We conclude that no simplifications of equations~\eqref{eq:br} and \eqref{eq:bphi} or,
equivalently, \eqref{eq:blf_1} and \eqref{eq:blg_1} are either justifiable or expedient
in the case of quadrupolar solutions. As we argue in section~\ref{Disc}, the
dipolar asymptotics for $|D|\gg1$ are of limited usefulness in a thin disc because the dipolar
modes are excited at dynamo numbers of so large a magnitude that the spatial scale of the
solution no longer exceeds the turbulent scale.

\bibliographystyle{gGAF}
\bibliography{asympt}

\label{lastpage}
\end{document}